\documentclass[aps,pra,twocolumn,superscriptaddress,tightenlines,floatfix]{revtex4-2}

\usepackage{amsmath,amssymb}
\usepackage{dsfont}
\usepackage{graphicx} 

\usepackage{hyperref}   %
\usepackage{placeins}   %
\usepackage{accents}    %
\usepackage{siunitx}    %
\usepackage{xcolor}     %
\graphicspath{{../},{figures/}}

\DeclareMathOperator*{\id}{\hat{1}}

\newcommand{\sket}[1]{\left| #1 \right\rangle\!\rangle}

\newlength{\dhatheight}
\newcommand{\dhat}[1]{%
    \settoheight{\dhatheight}{\ensuremath{\hat{#1}}}%
    \addtolength{\dhatheight}{-0.35ex}%
    \hat{\vphantom{\rule{1pt}{\dhatheight}}%
    \smash{\hat{#1}}}}
\def\comm#1#2{{\left[ #1,#2 \right]}}
\def\acomm#1#2{{\left\{ #1,#2 \right\}}}

\def\dv#1#2{{\frac{{\rm d}#1}{{\rm d}#2}}}
\def\pdv#1#2{{\frac{\partial }{\partial #2} #1 }}

\def\matrix#1{{\begin{pmatrix} #1 \end{pmatrix}}}
\def\mel#1#2#3{{\langle #1| #2 |#3\rangle}}
\def\dyad#1#2{{|#1\rangle \langle #2 |}}
\def\abs#1{{|#1|}}
\def\ket#1{{|#1\rangle}}
\def\bra#1{{\langle #1|}}
\def\Tr#1{{\rm Tr}\!\left( #1 \right)}

\newcommand{\NHH}{\hat{H}_{_{\rm {NH}}}}
\newcommand{\PT}{${\mathcal{PT}}$}

\def \etal{\textit{et al.}}

\begin{document}

\title{Liouvillian and Hamiltonian exceptional points of atomic
vapors:\\ The spectral signatures of quantum jumps}

\author{Marek Kopciuch}
\email[]{marek.kopciuch@amu.edu.pl} \affiliation{Institute of
Spintronics and Quantum Information, Faculty of Physics and
Astronomy, Adam Mickiewicz University, 61-614 Pozna\'{n}, Poland}

\author{Adam Miranowicz}
\email[]{adam.miranowicz@amu.edu.pl} \affiliation{Institute of
Spintronics and Quantum Information, Faculty of Physics and
Astronomy, Adam Mickiewicz University, 61-614 Pozna\'{n}, Poland}

\date{\today}

\begin{abstract}
We investigate spectral singularities in an alkali-metal atomic
vapor modeled using four and effectively three hyperfine states.
By comparing the eigenvalue spectra of a non-Hermitian Hamiltonian
(NHH) and a Liouvillian superoperator, we analyze the emergence
and characteristics of both semiclassical and quantum exceptional
points. Our results reveal that, for atomic systems, the NHH
approach alone may be insufficient to fully capture the system's
spectral properties. While NHHs can yield accurate predictions in
certain regimes, a comprehensive description typically requires
the Liouvillian formalism, which governs the Lindblad master
equation and explicitly incorporates quantum jump processes
responsible for repopulation dynamics. We demonstrate that the
inclusion of quantum jumps fundamentally alters the spectral
structure of the system. In particular, we present examples in
which the existence, location in parameter space, or even the
order of spectral degeneracies differ significantly between the
two approaches, thereby highlighting the impact of quantum jumps
and the limitations of the NHH method. Finally, using the
hybrid-Liouvillian formalism, we show how quantum jumps reshape
spectral features initially predicted by the NHH, ultimately
determining the full Liouvillian spectrum.
\end{abstract}

\maketitle

\section{Introduction}
\label{sec:introduction}

Non-Hermitian quantum physics has recently attracted considerable
attention for its ability to describe energy dissipation in open
quantum systems and to predict exotic phenomena such as EPs--non-Hermitian degeneracies first studied
in the 1960s~\cite{KatoBook}. However, intense research into EPs
began in the early 2000s~\cite{Heiss2000, Dembowski2001,
Berry2004}, spurred in part by the introduction of
$\mathcal{PT}$-symmetric quantum mechanics based on NHHs~\cite{Bender1998, Bender2007}.

Numerous studies investigating NHHs and their singularities, often
termed Hamiltonian EPs (HEPs), in engineered non-Hermitian systems
have emerged across a wide range of fields as reviewed
in~\cite{Heiss2012, Ozdemir2019, Miri2019, Parto2021}). These
include: optics~\cite{Guo2009, Ruter2010, Regensburger2012,
Zhen2015, Ding2015, Cerjan2016, Feng2017, El-Ganainy2019,
Parto2021}, electronics~\cite{Schindler2011},
plasmonics~\cite{Benisty2011, Alaeian2014, Kuo2020},
acoustics~\cite{Zhu2014, Jing2014, Fleury2015, Ding2016, Lu2017,
Zhang2018}, cavity optomechanics~\cite{Jing2014, Jing2015,
Schonleber2016, Xu2016, Jing2017}, atom optics~\cite{Zhang2016},
circuit quantum electrodynamics (QED)~\cite{Naghiloo2019,
Chen2021, Chen2022}, and cavity QED~\cite{Roy2021a, Roy2021b,
Jahani2021}.  A variety of physical platforms have been employed
in these investigations, such as photonic~\cite{Regensburger2012,
Zhen2015, Ding2015, Cerjan2016, Sheikhey2021} and
atomic~\cite{Zhang2016} lattices, metamaterials~\cite{Kang2013,
Fleury2014, Sun2014, Kang2016, Xiao2016}, exciton-polariton
systems~\cite{Gao2015}, atomic vapors~\cite{Peng2016}, trapped
ions~\cite{Ding2021}, or thermal atomic
ensembles~\cite{Liang2023}. Numerous proposed applications of EPs
span a wide range of fields, including~\cite{Ozdemir2019,
Miri2019}: quantum control and state engineering, quantum
thermodynamics, mode conversion and switching, topological energy
transfer, dynamic stability control, non-Hermitian quantum
information processing, as well as neuromorphic and reservoir
computing. Considerable attention has been directed toward signal
amplification and spectral filtering near EPs, particularly for
their potential use in exceptional-point-enhanced quantum
sensing~\cite{Miller2017, Wiersig2020a, Wiersig2020b}.

Effective NHHs can accurately describe coherent, nonunitary
dynamics in classical and semiclassical systems. However, they
fall short in capturing the full behavior of quantum systems
subject to quantum jumps and associated noise corresponding to the
loss or gain of an excitations (like photons, phonons, or magnons)
exchanged with the environment, which monitors (measures) the
system.  A fully quantum treatment is therefore
required---typically formulated through a Liouvillian
superoperator derived from a master equation, or equivalently, via
Fokker-Planck or Heisenberg-Langevin equations.

To address the limitations of HEPs derived from NHHs, the concept
of quantum Liouvillian exceptional points (LEPs) was
introduced~\cite{Hatano2019, Minganti2019}. LEPs are defined as
spectral degeneracies of Liouvillian superoperators, where both
eigenvalues and eigenvectors coalesce, paralleling the notion of
HEPs in NHHs. By incorporating quantum jumps explicitly, LEPs
extend the applicability of the HEP framework, enabling a
consistent and physically complete description of decoherence and
noise in open quantum systems, while preserving the structure of
canonical commutation relations.

For classical or semiclassical systems---where quantum jumps are
negligible---HEPs and LEPs yield consistent predictions. However,
in fully quantum systems, the presence of quantum jumps can
significantly modify the structure of EPs: they may generate,
shift, or even eliminate EPs, or reduce their order, as shown in
this work. It should be emphasized that quantum jumps (responsible
for repopulation processes) are particularly important in the
context of atomic ensembles, as they underlie one of the most
widely used techniques in atomic physics, i.e., optical pumping
\cite{AuzinshBook}.

The connection between HEPs and LEPs can be naturally
elucidated---and even continuously interpolated---within the
hybrid Liouvillian formalism of Ref.~\cite{Minganti2020}, by
post-selecting experimental outcomes based on the number of
quantum jumps. The two limiting cases correspond to lack of
quantum jumps (recovering HEPs) and an arbitrary number of jumps
(leading to LEPs).

It should be stressed that LEPs are embedded within the
well-established formalism of open quantum systems, making their
interpretation and computation more transparent and physically
grounded. Unlike NHH-based approaches, which often require the
computation of system-specific metrics to ensure physical
validity---following the formalism of \PT-symmetric quantum
mechanics~\cite{Bender1998, Bender2007, Ju2019}---the analysis of
Liouvillians and their exceptional points relies solely on
standard quantum mechanics. This eliminates the ambiguities and
potential pitfalls of the NHH framework, which, although
mathematically rich and conceptually intriguing~\cite{Ju2022,
Ju2024, Ju2025}, can lead to misleading or incorrect conclusions
(like apparent violations of no-go quantum information
theorems~\cite{Ju2019}) if applied imprecisely.

Until recently, most research on EPs has focused on HEPs in
classical or semi-classical systems, but interest in LEPs of fully
quantum systems has been steadily increasing after their
introduction in 2019---particularly driven by recent experimental
advances in circuit QED~\cite{Chen2021, Chen2022, Erdamar2024,
Abo2024} and trapped ion platforms~\cite{Zhang2022, Bu2023,
Bu2024}. Proposed applications of LEPs include enhanced quantum
sensing and precise dynamical control of quantum systems. Notably,
LEPs play a significant role in quantum thermodynamics---for
instance, they act as indicators of critical decay regimes in
quantum thermal machines~\cite{Khandelwal2021}, and experimental
findings suggest they may enhance the efficiency of quantum heat
engines~\cite{Bu2023, Erdamar2024}. Nevertheless, a more
comprehensive investigation of the role of LEPs in comparison to
HEPs is needed to fully exploit their potential for emerging
quantum technologies.

LEPs have also been proposed and studied in the context of
non-Markovian processes~\cite{Lin2025}. However, the present work
focuses on dissipative dynamics and the corresponding LEPs within
the framework of the standard Lindblad master equation.

Moreover we study only EPs, but it would be interesting to search
for other types spectral degeneracies including diabolical points
(defined by degenerate eigenvalues with the corresponding
orthogonal eigenvectors)~\cite{Berry1984} and hybrid points, which
are higher-order degeneracies combining properties of exceptional
and diabolical points~\cite{Arkhipov2023, Naikoo2025}. It is also
worth noting that quantum exceptional, diabolical, and hybrid
degeneracies can alternatively be characterized using the
Heisenberg-Langevin formalism, as demonstrated in
Refs.~\cite{Perina2022, Perina2023, Thapliyal2024, Perina2024}.

In this work, motivated by the experimental observation of HEPs in
alkali-metal atomic vapors modeled with a few hyperfine
states~\cite{Liang2023}, we investigate the emergence of LEPs
within the same framework---including its generalization to
account for detunings---and compare these quantum LEPs with the
corresponding HEPs predicted by the non-Hermitian Hamiltonian
approach.

Alkali-metal atomic vapors represent one of the most versatile
platforms in atomic optics and quantum technologies. These systems
operate across a wide range of conditions, enabling the study of
diverse quantum phenomena. In the ultracold regime, alkali atoms
are essential for realizing Bose-Einstein
condensates~\cite{Anderson1995, Ketterle1995, Greiner2002} and
optical lattices~\cite{Greiner2002, Jessen1996}, which enable
implementations of various quantum information algorithms. In the
opposite thermal limit, known as the hot-vapor regime,
temperatures often exceed $100^\circ$C, allowing operation in the
spin-exchange relaxation-free (SERF) regime~\cite{Happer1977,
Mouloudakis2024, Kong2020}. In this regime, rapid interatomic
collisions suppress decoherence, enabling collective spin
dynamics~\cite{Mouloudakis2024}, extended coherence times, and the
observation of many-body entanglement~\cite{Kong2020}.

Lying between these ultracold and hot vapor regimes is the
room-temperature domain, where alkali vapors confined in
anti-relaxation-coated cells (e.g., paraffin-coated) exhibit
surprisingly long spin coherence times~\cite{Balabas2010}. This
regime has enabled the observation of a wide range of quantum
effects, including coherent population
trapping~\cite{Schmidt1996}, spin squeezing~\cite{Hammer2004},
macroscopic entanglement~\cite{Katz2020, Jensen2011}, spin wave
dynamics~\cite{Jensen2011}, and the generation of squeezed and
entangled light modes~\cite{Hammer2004, Mikhailov2008}.

Remarkably, the room-temperature and SERF regimes have also
provided the foundation for developing ultra-sensitive quantum
sensors. These include magnetometers~\cite{Dang2010,
Chalupczak2012, Budker2007}, atomic gyroscopes~\cite{Budker2007,
Kornack2005}, and detectors for exotic physics, such as searches
for dark matter~\cite{Afach2021, Padniuk2022}. The combination of
long coherence times, collective quantum behavior, and accessible
experimental setups makes alkali-metal vapors a powerful testbed
for investigating EPs, particularly in the context of enhanced
quantum sensing.

Specifically, we employ an effective four-level model (and its
effective three-level version) to describe optical transitions
between two hyperfine manifolds in atomic vapors, characterized by
total angular momenta $f$ (ground state) and $F$ (excited state).
Of particular interest is the minimal configuration that still
captures the essential dynamical features---namely, the transition
$f = 1 \rightarrow F = 0$. This simple model is sufficient to
distinguish the interactions of atomic states with light of
different polarizations.

To ensure that the model faithfully represents real atomic
transitions, it is crucial to account for the characteristic
timescales of the system. In typical experimental realizations,
the excited-state relaxation time is several orders of magnitude
shorter than the timescales governing ground-state evolution. This
separation of timescales justifies a common approximation:
eliminating the excited state to obtain an effective description
of the ground-state dynamics.

While adiabatic elimination and direct computation under the
assumption of negligible excited-state population are standard
techniques for this reduction, in this work we adopt a more
systematic approach. Specifically, we utilize the effective
operator formalism for open quantum systems, as introduced in
Ref.~\cite{Reiter2012} and detailed in
Sec.~\ref{sec:model_eff_Hamiltonian}. This method allows for a
controlled dimensionality reduction and is especially advantageous
for systems with larger excited-state angular momentum.

The paper is organized as follows. In
Sec.~\ref{sec:introduction_EP}, we review the essential formalisms
and definitions of EPs associated with NHHs, quantum Liouvillians,
and hybrid Liouvillians. Section~\ref{sec:model_eff_Hamiltonian}
introduces an effective, description of the slowly-varying
ground-state dynamics in our system. Section~\ref{sec:no_RF_case}
discusses the model without radio-frequency (RF) detunings in the context of both
the NHHs and Liouvillian superoperators. In
Sec.~\ref{sec:RF_case}, we examine a generalized model with an
additional detuning of the RF magnetic field, in which the HEPs
and LEPs exhibit significant differences. Technical details,
including lengthy derivations and formulas, are provided in the
Appendix for the reader's convenience. We present our
conclusions in Sec.~\ref{sec:conclusions}.

\section{ Quantum and semiclassical exceptional points:
Basic concepts} \label{sec:introduction_EP}

A key objective of this article is to compute and compare the
differences between EPs arising from the
non-Hermitian part of the system's Hamiltonian---which governs
coherent yet dissipative dynamics---and those of the full
Liouvillian, which additionally incorporates quantum jumps.
Following Ref.~\cite{Minganti2019}, these Hamiltonian and
Liouvillian EPs are commonly referred to as (semi)classical and quantum
EPs, respectively. To this end, we begin by recalling the basic
definitions.

Exceptional points are singularities in a system's parameter space
where two or more eigenvalues, along with their corresponding
eigenvectors, coalesce. Unlike conventional degeneracies, EPs
involve not only eigenvalue coincidence but also a collapse of the
eigenspace, resulting in linearly dependent eigenvectors.

EPs are a hallmark of non-Hermitian systems---i.e., systems
exhibiting loss, gain, or other non-unitary dynamics. As such,
their analysis requires consideration of either an NHH
spectrum or the full Liouvillian superoperator via the
Lindblad master equation:
\begin{equation}
\dot\rho = \mathcal{L}(\rho) = -i \comm{\hat{H}}{\rho} -
\sum_{\mu} \left( \frac{1}{2} \acomm{\hat{L}_{\mu}^{\dagger}
\hat{L}_{\mu}}{\rho} - \hat{L}_{\mu} \rho \hat{L}_{\mu}^{\dagger}
\right), \label{eq:03_master_equation}
\end{equation}
where $\mathcal{L}$ represents the Liouvillian superoperator
acting on the density matrix, and $\hat{L}_\mu$ denotes quantum
jump operators. The key distinction between these two approaches
lies in the treatment of quantum jumps: while NHHs account for
dissipative evolution, only the Liouvillian framework captures the
stochastic repopulation effects associated with quantum jumps.
Equation~(\ref{eq:03_master_equation}) can be rewritten by
defining an effective NHH:
\begin{equation}
    \NHH = \hat{H} - \frac{i}{2} \sum_{\mu} \hat{L}_{\mu}^{\dagger} \hat{L}_{\mu},
    \label{eq:03_non_hermitian_ham}
\end{equation}
which leads to the following transformation
\begin{equation}
\begin{split}
- i \comm{\NHH}{\rho}
 &\equiv -i \big(\NHH\rho-\rho \NHH^\dagger \big) \\
 &= \comm{\hat{H}}{\rho} - \frac{1}{2}\sum_{\mu}
\acomm{\hat{L}_{\mu}^{\dagger} \hat{L}_{\mu}}{\rho}.
\label{eq:03_HNH_commutator}
\end{split}
\end{equation}
It is important to note that the commutator here includes complex
conjugation, which is typically omitted when $\hat{H}$ is a
self-adjoint operator. This generalized formulation allows us to
express the master equation as
\begin{equation}
    \dv{\rho}{t} = \mathcal{L}(\rho)
    = -i \comm{\NHH}{\rho} + \sum_{\mu} \hat{L}_{\mu} \rho \hat{L}_{\mu}^{\dagger}.
\end{equation}
This expression underscores the fundamental difference between
analyzing LEPs, which account for
the complete quantum dynamics including quantum jumps, and
HEPs, which reflect only the
coherent yet dissipative evolution described by the NHH.

Even though these two approaches may seem difficult to
compare---since there is an apparent discontinuity due to the
presence or absence of quantum jumps---there exists a way to
interpolate this transition. This can be achieved by following the
hybrid Liouvillian formalism~\cite{Minganti2020}:
\begin{equation}
\begin{split}
    \mathcal{L}'(\rho) &= -i \comm{\hat{H}}{\rho} - \sum_{\mu} \left( \frac{1}{2} \acomm{\hat{L}_{\mu}^{\dagger} \hat{L}_{\mu}}{\rho} - q \hat{L}_{\mu} \rho \hat{L}_{\mu}^{\dagger}  \right)\\  &= -i \comm{\NHH}{\rho} + q \sum_{\mu} \hat{L}_{\mu} \rho \hat{L}_{\mu}^{\dagger},
    \label{eq:03_hybrid_liouvillian}
\end{split}
\end{equation}
where $q$ is the quantum-jump parameter. In this framework, one
possible interpretation is that it corresponds to an experiment
with postselection, where only trajectories without quantum jumps
($q=0$) are considered. In this limit, the evolution is governed
purely by the NHH. More generally, the hybrid Liouvillian can be
understood as describing an experiment in which the selection
process leading to postselection is imperfect. By continuously
varying $q$ from 0 to 1, one obtains a smooth and physically
meaningful interpolation between the spectrum of the NHH and the
full Liouvillian spectrum.


It is worth noting that computing the spectrum of a
Hamiltonian---whether Hermitian or non-Hermitian---is relatively
straightforward, as it involves diagonalizing a matrix whose
dimension corresponds to that of the system's Hilbert space. In
contrast, analyzing the spectrum of the Liouvillian superoperator
poses additional challenges. As shown in
Eq.~\eqref{eq:03_master_equation}, the superoperator $\mathcal{L}$
acts on the density matrix in a nontrivial way, rather than as a
simple matrix-vector multiplication. Nevertheless, because
$\mathcal{L}$ is a linear, completely positive, and
trace-preserving map, it can be recast as a matrix acting on the
vectorized space of operators---that is, as a linear operator on
the Liouville space. This representation allows the use of
standard linear algebra techniques, albeit in a space whose
dimension is the square of the Hilbert space dimension.

Importantly, while the spectrum of an NHH can be computed using
standard diagonalization techniques, it is often advantageous to
express the Hamiltonian in the superoperator formalism. This
facilitates a direct comparison with the Liouvillian spectrum and
provides a systematic way to identify how quantum jumps influence
the system's dynamics.

Since the Liouvillian is a superoperator---acting on operators
rather than vectors---it can be viewed as an operator on an
extended space, where elements are vectorized representations of
$d \times d$ matrices. Here, $d$ denotes the dimension of the
system's Hilbert space. Crucially, the action of the Liouvillian
is not restricted to physical density matrices, which are
Hermitian, positive semi-definite, and trace-one. As a result, its
eigenvectors often correspond to matrices that do not represent
valid physical states---they may be non-Hermitian or have non-unit
trace. Nevertheless, these eigenvectors of a nonsingular
Liouvillian (so except LEPs) span the space of all $d \times d$
matrices, forming a complete basis. This enables one to decompose
any physical density matrix in terms of these eigenvectors, with
each component evolving independently according to the
corresponding eigenvalue of the Liouvillian. This decomposition
mirrors the role played by the Hamiltonian eigenbasis in the
unitary evolution of closed quantum systems, offering a powerful
tool for analyzing open-system dynamics.

One can classify the behavior of Liouvillian eigenvectors based on
their associated eigenvalues as follows (see, e.g.,
Refs.~\cite{Baumgartner2008, Albert2014, Abo2024}): (1) If an
eigenvalue of the Liouvillian is zero, the corresponding
eigenvector represents a stationary state of the system, remaining
unchanged under the Liouvillian evolution. (2) If an eigenvalue
is purely imaginary, the corresponding eigenvector undergoes
periodic, undamped evolution. (3) If an eigenvalue is real and
negative, the corresponding eigenvector exhibits simple
exponential damping. (4) If an eigenvalue is complex with a
negative real part, the corresponding eigenvector undergoes damped
oscillatory evolution, where the imaginary part determines the
oscillation frequency and the real part sets the decay rate.

\section{Effective Non-Hermitian Hamiltonian for the Model}
\label{sec:model_eff_Hamiltonian}

We analyze the theoretical model presented in
Ref.~\cite{Liang2023}. The system consists of a ground state with
total angular momentum $f=1$ and an excited state with $F=0$. The
system is subject to two magnetic fields: a static leading field
along the $\Vec{z}$ axis, characterized by an effective Larmor
frequency $\Omega_L$, and a perpendicular oscillating field of the
form $B_{\rm{RF}} \cos\left(\omega_{\rm{RF}} t\right) \Vec{x}$,
which induces coupling with an effective RF Rabi frequency $J =
\gamma_B B_{\rm{RF}}/\sqrt{2}$, where $\gamma_B$ is the
gyromagnetic ratio of the given state. Additionally, the system
interacts with linearly polarized light with polarization $\pi$.
This results in the following Hamiltonian:
\begin{equation}
    \hat{H}_0 =
    \matrix{ \Omega_L & J c_t &0&0\\
    J c_t &0&J c_t &-\Omega_R \cos(\omega t)\\
    0&J c_t&-\Omega_L&0\\
    0&-\Omega_R \cos(\omega t)&0&\omega_0
    },
    \label{eq:03_time_dep_hamiltonian}
\end{equation}
where $c_t=\cos(\omega_{\rm{RF}}t)$, $\omega$ is the laser frequency, $\omega_0$ is the transition frequency between the ground and excited states, and $\Omega_R$ is the Rabi frequency of the optical transition.

\begin{figure}[t]
    \centering
    \includegraphics[width=\columnwidth]{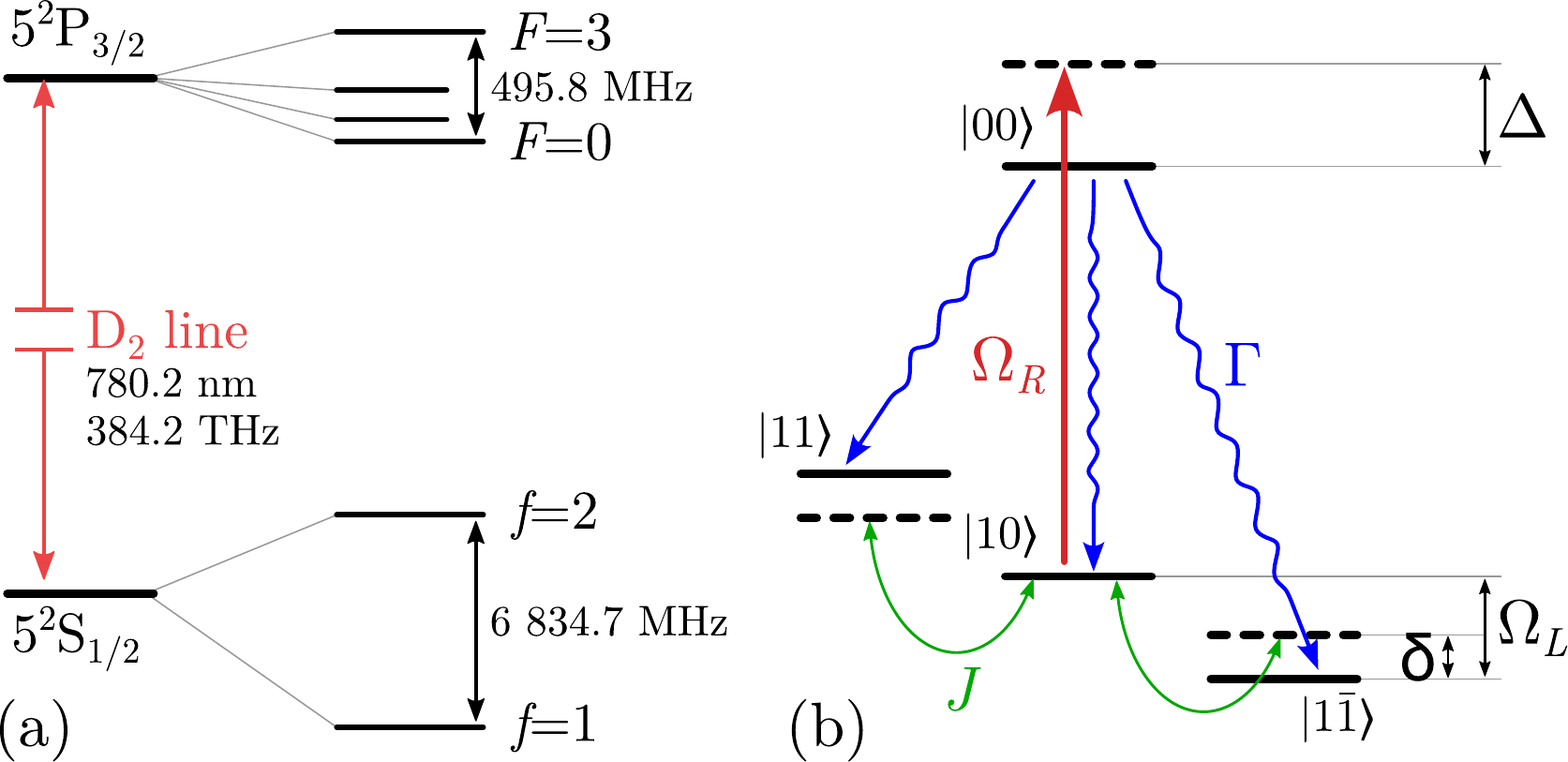}
    \caption{
(a) Energy-level diagram of the D$_2$ transition in ${}^{87}$Rb,
relevant for a potential experimental implementation of the
studied system due to the presence of the $f = 1 \rightarrow F =
0$ transition. (b) Schematic of the relevant subspace spanned by
the Zeeman sublevels. The red arrow indicates the optical
transition, driven by a field with Rabi frequency $\Omega_R$ and
detuning $\Delta$. Blue wavy arrows represent spontaneous emission
channels associated with different photon polarizations, occurring
at rate $\Gamma$. Green arrows depict radio-frequency (RF)
magnetic coupling of strength $J$, detuned by $\delta$ from the
Larmor frequency $\Omega_L$.}
    \label{fig:energy_level_diagram}
\end{figure}

To eliminate the time dependence from the Hamiltonian, the authors
of Ref.~\cite{Liang2023} first apply the standard optical rotating
wave approximation (RWA), followed by a secondary transformation
at the magnetic Larmor frequency associated with the ground state.
Alternatively, a generalized RWA can be used to obtain a compact form of this transformation (for
details see Appendix~\ref{app:A3}). The transformed Hamiltonian
takes the form:
\begin{equation}
    \hat{H} =
    \matrix{-\delta&J&0&0 \\ J&0&J&-\Omega_R \\ 0&J&\delta&0 \\ 0&-\Omega_R&0&-\Delta},
    \label{eq:03_time_indep_hamiltonian}
\end{equation}
where $\delta = \omega_{\rm{RF}} - \Omega_L$, $\Delta = \omega -
\omega_0$, and $\omega_0$ is the excited state energy. For
convenience, we adopt the rescaling $J/2 \rightarrow J$ and
$\Omega_R/2 \rightarrow \Omega_R$.


To simplify the analysis of the four-level system, we reduce it to
an effective three-level system that captures only the slow
ground-state dynamics. This reduction is performed using the
effective operator formalism, which was introduced in Ref.~\cite{Reiter2012}, while the detailed calculations are provided in
Appendix~\ref{app:A_effective_system_description}. The essential
components of the Hamiltonian relevant to this effective
description are identified as follows:
\begin{eqnarray}
    \hat{H}_g &=& -\delta \hat{F}_z + J \sqrt{2} \hat{F}_x,\\
    \hat{V}_+ &=& \left( \hat{V}_- \right)^{\dagger} = -\Omega_R \dyad{00}{10},\\
    \hat{H}_e &=& -\Delta \dyad{00}{00},
\end{eqnarray}
where $\hat{H}_{g(e)}$ is the ground (excited) state Hamiltonian,
and $\hat{F}_i$ is the angular momentum operator along the $i$
axis. Next, the spontaneous-emission Lindblad operator for
polarization $\varepsilon$, as defined in
Eq.~\eqref{eq:03_Lindblad}, takes the following form:
\begin{equation}
    \hat{L}^{\rm{sp}}_{\varepsilon} = i \sqrt{\dfrac{\Gamma}{3}} \dyad{1\bar{\varepsilon}}{00}.
    \label{eq:spont_emission}
\end{equation}
Using Eqs.~\eqref{eq:03_excited_nh_ham}, \eqref{eq:03_relax_q},
and \eqref{eq:03_relax_tot}, we obtain the NHH for the excited state and its inverse:
\begin{equation}
    \begin{split}
    \hat{H}_{\rm{eNH}} &=- \dfrac{2\Delta + i \Gamma}{2}
    \dyad{00}{00},\\
    \hat{H}_{\rm{eNH}}^{-1} &=-\dfrac{4\Delta- 2i\Gamma}
    {\Gamma^2 + 4\Delta^2} \dyad{00}{00}.
    \end{split}
\end{equation}
From Eqs.~\eqref{eq:03_eff_ham} and \eqref{eq:03_eff_Lindblad}, it
follows that
\begin{eqnarray}
    \hat{H}_{\rm{eff}} &=& \hat{H}_g + \dfrac{4\Delta\Omega_R}
    {\Gamma^2+4\Delta^2} \dyad{10}{10},\\
    \hat{L}_{\rm{eff}}^{\varepsilon} &=& \dfrac{2\sqrt{\Gamma}\Omega_R}{\sqrt{3}
    (\Gamma - 2i\Delta)}\dyad{1\bar{\varepsilon}}{10}.
\end{eqnarray}
Substituting these results into
Eq.~\eqref{eq:03_non_hermitian_ham}, we obtain the effective NHH:
\begin{equation}
\begin{split}
    \NHH &= \hat{H}_g - \dfrac{2i \Omega^2_R}{\Gamma-2i\Delta} \dyad{10}{10}\\ &=
    \matrix{-\delta&J&0 \\ J&- \dfrac{2i \Omega^2_R}{\Gamma-2i\Delta}&J \\ 0&J&\delta}.
\end{split}
\label{eq:03_eff_ham_article}
\end{equation}
Finally, by assuming the light is resonant with the transition
($\Delta \to 0$), we recover Eq.~(2) from Ref.~\cite{Liang2023}.


\begin{figure}[h!]
    \centering
    \includegraphics[width=\columnwidth]{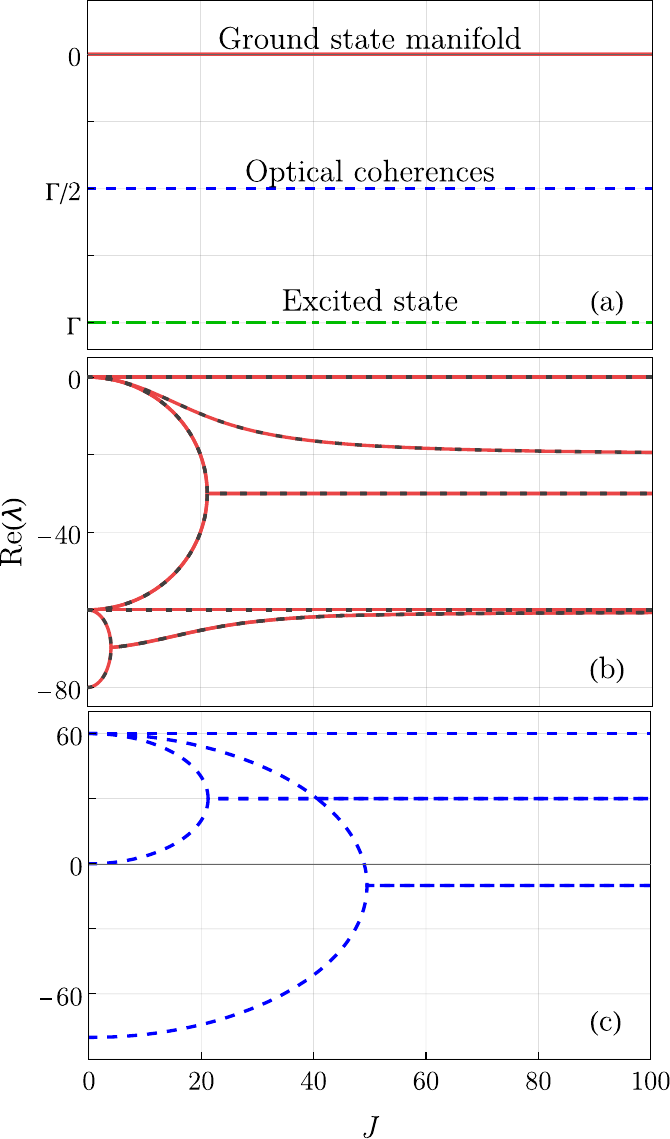}
    \caption{
Real part of the Liouvillian spectrum for the full four-level
atomic system as a function of the coupling strength $J$
calculated from Eq.~\eqref{eq:03_master_equation} with
Eqs.~\eqref{eq:03_time_indep_hamiltonian},
and~\eqref{eq:spont_emission}. Panel (a) presents the complete
spectrum, while panels (b) and (c) provide magnified views of two
spectrally distinct regions. These regions are identified based on
their separation by characteristic spectral gaps: (b) corresponds
to the ground-state manifold, and (c) to optical coherences. The
excited-state contribution, consisting of a single level, is not
magnified but remains visible in panel (a) as a green dash-dotted
line. The color coding in panels (b) and (c) is consistent with
that used in panel (a). The only distinction is that
black dashed lines in panel (b) represent the
spectrum of the effective ground-state Liouvillian as described by
Eq.~\eqref{eq:hybrid_Liouvillian_matrix}, illustrating the
excellent agreement with the full system, with deviations not
visible at the displayed scale. Note that the horizontal axis is
shifted by \(\Gamma/2\) in (b) and by \(\Gamma\) in (c) to better
center the respective spectral regions. Assumed parameters:
\(\Gamma = 2\pi \times 5.7 \times 10^6\), \(\Omega = 30\).}
    \label{fig:spec_full_parts}
\end{figure}

\section{RF Tuned Regime of the Model}
\label{sec:no_RF_case}

Here, we present both analytical and numerical analyses of the
simplified NHH model described by
Eq.\eqref{eq:03_eff_ham_article}, assuming that the RF magnetic
field is tuned to the Larmor frequency ($\delta = 0$). In the case
of the NHH alone, our results are consistent with the reasoning
presented in Ref.~\cite{Liang2023}. However, the inclusion of
quantum jumps significantly alters the spectral properties, even
in this simplified scenario.

\subsection{Validation of the effective three-level model}
\label{sec:no_RF_case_validation}

At the beginning, we slightly depart from chronological order by
first analyzing the Liouvillian of the full four-level system,
obtained using the Liouville-Fock basis (see
Appendix~\ref{app:B2}), as well as that of the effective
three-level system described by
Eq.~\eqref{eq:hybrid_Liouvillian_matrix}. This analysis is
intended to demonstrate that the subsequent use of the effective
model is fully justified within the parameter regime relevant for
atomic vapor systems.

To rigorously compare the spectra, we performed numerical
simulations using system parameters chosen within experimentally
accessible ranges. In all cases, the spectra are governed by two
parameters: the coupling strength of the oscillatory magnetic
field (i.e., the magnetic field Rabi frequency, $J$), and the
reduced optical Rabi frequency -- $\Omega$. The parameter $\Omega$
can be tuned by varying the coupling light intensity within the
range from \SI{1}{\micro\watt} to \SI{10}{\milli\watt}, allowing for
values in the range $10 < \Omega < 1000$.

It is important to note that the Liouvillian superoperator
$\dhat{\mathcal{L}}$ depends on an additional parameter, $\Gamma$,
which represents the natural linewidth (i.e., relaxation rate) of
the excited state. For the experimentally employed $D_2$
transition (see the $F=1 \rightarrow F=0$ transition shown in
Fig.~\ref{fig:energy_level_diagram}) of $^{87}$Rb, this parameter
is given by $\Gamma = 2\pi \times \SI{5.746}{\mega\hertz}$. One
might also consider the $F=1 \rightarrow F=1$ transition present
in the $D_1$ line, where $\Gamma = 2\pi \times
\SI{6.065}{\mega\hertz}$. However, as shown in the figures, this
difference does not qualitatively affect the spectrum and has
negligible quantitative impact on the part of the spectrum
corresponding to the effective system.

An examination of Fig.~\ref{fig:spec_full_parts}(a) reveals a
clear grouping of eigenvalues, which can be classified as
$\lambda_i \approx 0$, $\Gamma/2$, or $\Gamma$. Specifically, the
first group contains nine eigenvalues, the second group six, and
the third group a single eigenvalue. This classification enables a
straightforward physical interpretation: The first group
corresponds to ground-state evolution, the second to
ground-excited-state optical coherences, and the third to the
excited state itself.

This also explains why small variations in $\Gamma$ have little
effect on the spectrum: the energy separation between these groups
is large compared to the internal dynamics within each group,
rendering the system robust to minor changes in $\Gamma$.

\subsection{Spectrum of the non-Hermitian Hamiltonian operator}
\label{sec:no_RF_case_operator}

Following the approach in Ref.~\cite{Liang2023}, one can
demonstrate that by setting both the optical field detuning
$\Delta$ and the oscillating magnetic field detuning $\delta$ to
zero, while also introducing the reduced Rabi frequency $\Omega =
\Omega_R^2/\Gamma$, the Hamiltonian from
Eq.~\eqref{eq:03_eff_ham_article} can be expressed as:
\begin{equation}
    \NHH = \matrix{0&J&0 \\ J&-2i \Omega&J \\ 0&J&0 }.
    \label{eq:03_Hnh_no_relax}
\end{equation}
Its eigenvalues are readily obtained as:
\begin{equation}
    E_0 = 0, \qquad E_{\pm} = -i\Omega \pm \sqrt{2J^2 - \Omega^2},
\end{equation}
with the corresponding eigenvectors given by:
\begin{eqnarray}
    \ket{E_0} &=& -\ket{11} + \ket{1\bar{1}},\\
    \ket{E_{\pm}} &=& \ket{11}-\dfrac{i \Omega \pm \sqrt{2J^2 - \Omega^2}}{J} \ket{10} + \ket{1\bar{1}}.
    \label{eq:03_eigenvect}
\end{eqnarray}
From this expression, it becomes evident that when $2J^2 = \Omega^2$, the system exhibits a second-order HEP, characterized by the coalescence of both eigenvalues $E_+$ and $E_-$, as well as their corresponding eigenvectors.
The reduced Rabi frequency was introduced to simplify the
calculations; however, it is worth noting that it has a
straightforward relation to, often used experimental parameter, the fine and hyperfine saturation
parameters: $\kappa_1 = \Omega/\Gamma$ and
$\kappa_2=\Omega/\gamma$.

\subsection{Additional isotropic relaxation}

It is important to note that, up to this point, hyperfine
isotropic relaxation has not been incorporated into our analysis.
This aspect was likewise omitted in Ref.~\cite{Liang2023}, where
the authors acknowledged its relevance but did not explicitly
include it in the NHH formulation. In our approach, this hyperfine
relaxation is introduced in the form:
\begin{equation}
    \hat{L}_{\mu}^{g} \rightarrow \hat{L}^{g}_{mn} = \sqrt{\dfrac{\gamma}{3}}
    \dyad{1m}{1n}, \quad m,n=\lbrace \pm1,0\rbrace,
\end{equation}
where $\gamma$ is the rate of ground-state isotropic relaxation.
This collection of relaxation channels results in isotropic
relaxation~\cite{Liang2023}. To confirm this, we evaluate the
corresponding effective relaxation operator (see
Appendix~\ref{app:A4})
\begin{equation}
\begin{split}
    \hat{\Gamma} &= \sum_{m,n} \left(\hat{L}^{g}_{mn}\right)^{\dagger}\hat{L}^{g}_{mn} \\ &= \dfrac{\gamma}{3} \sum_{m,n}
    \dyad{1n}{1m}\dyad{1m}{1n} = \gamma \id,
    \label{eq_Gamma}
\end{split}
\end{equation}
and the effective repopulation term (see Appendix~\ref{app:A4})
\begin{equation}
\begin{split}
    \hat{\Lambda}(\rho) &= \sum_{m,n} \hat{L}_{mn}^{g} \rho \left( \hat{L}_{m,n}^g \right)^{\dagger} \\ &= \dfrac{\gamma}{3} \sum_{m,n}
    \dyad{1m}{1n}\rho\dyad{1n}{1m} \\ &= \dfrac{\gamma}{3} \sum_{n} \rho_{nn} \id = \dfrac{\gamma}{3} \id.
    \label{eq_Lambda}
\end{split}
\end{equation}
By incorporating these relaxation effects into Eq.~\eqref{eq:03_eff_ham_article}, in accordance with Eq.~\eqref{eq:03_non_hermitian_ham} and under the previously stated assumptions, we arrive at the modified NHH:
\begin{equation}
    \NHH^{g} = \NHH - i \hat{\Gamma} =
    \matrix{-\dfrac{i \gamma}{2}&J&0 \\ J&-\dfrac{i}{2} (\gamma+4\Omega)&J \\ 0&J&-\dfrac{i \gamma}{2} }.
    \label{eq:03_Hnh_with_relax}
\end{equation}
This modification slightly alters the system's eigenenergies but
does not affect its eigenstates. The resultant eigenvalues are:
\begin{eqnarray}
    E_0 &=& -\dfrac{i \gamma}{2},\\
    E_{\pm} &=& -\dfrac{i}{2}(\gamma+2\Omega) \pm \sqrt{2J^2 - \Omega^2}.
\end{eqnarray}
Since such isotropic relaxation does not significantly alter the
spectral characteristics---producing only an overall isotropic
shift---we omit it in the subsequent analysis. (The details of how
isotropic relaxation modifies the superoperator spectrum are shown
in Appendix~\ref{app:C_isotropic}). Finally, we emphasize a subtle
but important difference between the relaxation model assumed in
our approach and that presented in the primary formulation of
Ref.~\cite{Liang2023}. Specifically, in our treatment, only the
ground state is subject to isotropic relaxation, whereas the
authors of Ref.~\cite{Liang2023} include relaxation of the excited
state as well. It is worth noting, however, that this distinction
is negligible for experimentally relevant parameters, since the
spontaneous emission rate $\Gamma$ typically exceeds the hyperfine
relaxation rate $\gamma$ by 4--6 orders of magnitude.

\begin{figure}[t]
    \centering
     \includegraphics[width=\columnwidth]{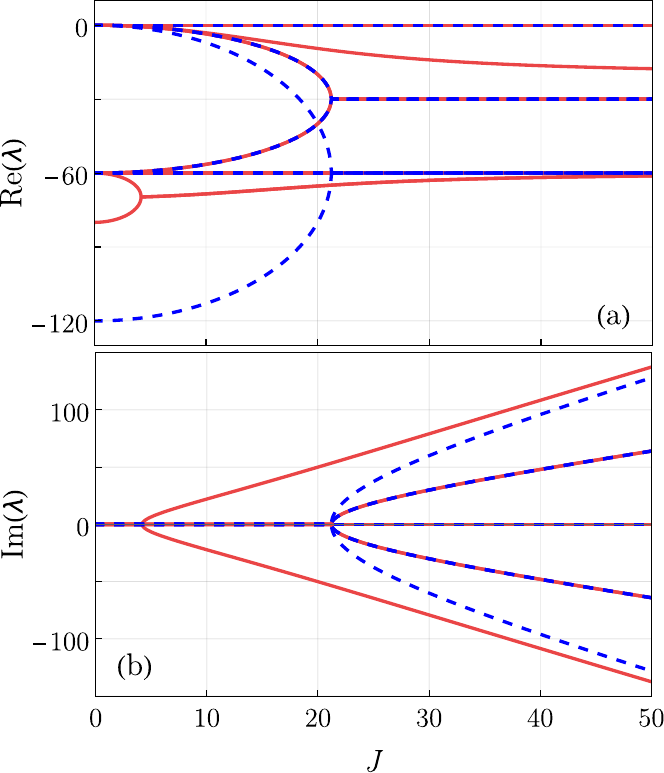}
\caption{Representative examples of the full spectra of the NHH
superoperator (blue dashed curves) compared with those of the
Liouvillian (red solid curves) as a function of $J$, obtained in
the limits $q=0$ and 1 of
Eq.~\eqref{eq:hybrid_Liouvillian_matrix}, respectively. Note that
the NHH superoperator corresponds to the hybrid Liouvillian in the
limit of zero quantum jump parameter ($q = 0$). Panels (a) and (b)
display the real and imaginary parts of the spectrum,
respectively. We set $\Omega = 30$.}
    \label{fig:tot_spec_k1}
\end{figure}

\subsection{Spectrum of the hybrid Liouvillian}
\label{sec:no_RF_case_superoperator}

In the next stage of analyzing the model without RF
detuning, we present calculations for the hybrid Liouvillian
model, which enables a fully quantum description of the system's
evolution by explicitly incorporating quantum jumps, as described
in Eq.~\eqref{eq:03_hybrid_liouvillian}. As a first step, we
neglect hyperfine relaxation (see Appendix~\ref{app:C_isotropic}), leading to:
\begin{equation}
    \dhat{\mathcal{L}}(q) = -i \dhat{H}_{\rm{NH}} + q\dhat{\Lambda}_{\rm{sp}},
\end{equation}
where $\dhat{\Lambda}_{\rm sp}$ represents the quantum jumps
induced by spontaneous emission. As a result, the Liouvillian
takes the following matrix form:
\begin{widetext}
\begin{equation}
    \dhat{\mathcal{L}}(q) = \left(
\begin{array}{ccccccccc}
 -2 \Omega  & 0 & 0 & 0 & J & 0 & 0 & 0 & 0 \\
 0 & -2 \Omega  & -2 J & -J & 0 & 0 & 0 & 0 & 0 \\
 0 & 2 J & -2 \Omega  & 0 & 0 & 0 & -J & \frac{2 \Omega }{\sqrt{3}} & \sqrt{\frac{8}{3}} \Omega  \\
 0 & J & 0 & 0 & 0 & 0 & -J & 0 & 0 \\
 -J & 0 & 0 & 0 & 0 & J & 0 & 0 & 0 \\
 0 & 0 & 0 & 0 & -J & -2 \Omega  & 0 & 0 & 0 \\
 0 & 0 & J & J & 0 & 0 & -2 \Omega  & -\sqrt{3} J & 0 \\
 0 & 0 & \frac{2 \Omega }{\sqrt{3}} & 0 & 0 & 0 & \sqrt{3} J & -\frac{2}{3} \Omega & -\frac{2}{3} \sqrt{2} \Omega  \\
 0 & 0 & -2 \sqrt{\frac{2}{3}} \Omega  (q-1) & 0 & 0 & 0 & 0 & \frac{2}{3} \sqrt{2} \Omega  (q-1) & \frac{4}{3} \Omega  (q-1) \\
\end{array}
\right).
\label{eq:hybrid_Liouvillian_matrix}
\end{equation}
\end{widetext}
Noting that the only distinction between the NHH and the
Liouvillian lies in the parameter $q$, it can be observed that
$\dhat{\mathcal{L}}$ closely resembles $\dhat{H}_{\rm NH}$. The
only difference appears in the last row. This is expected, as
the last row corresponds to the decay of the ninth component of
the vectorized density matrix, which is proportional to identity.
Since quantum jumps serve as a repopulation mechanism that ensures
probability conservation during the evolution, the vanishing last
row of the Liouvillian reflects this property.

The spectral analysis of this operator is particularly
interesting, as its limits at $q=0$ and $q=1$ correspond to the
spectra of the NHH superoperator and the Liouvillian,
respectively:
\begin{equation}
    \dhat{\mathcal{L}}(0) =  - i \dhat{H}_{\rm{NH}} \qquad
    \dhat{\mathcal{L}}(1) = \dhat{\mathcal{L}}
\end{equation}
We obtain the following spectrum of $\dhat{\mathcal{L}}'(q)$:
\begin{equation}
\begin{split}
&\left\{0,-2 \Omega ,-\alpha_1^{*},-\alpha_1^{*},-\alpha_{1},-\alpha_{1},\right.\\
&\left. \qquad -\frac{\phi }{\sqrt[3]{\zeta }}+\sqrt[3]{\zeta
}+\Omega  (2 q-9),\beta_+,\beta_-\right\},
\end{split}
\label{eq:hybrib_Liouvillian_spectra}
\end{equation}
where
\begin{equation}
  \beta_{\pm}=\frac{\phi}{\sqrt[3]{\zeta }} \frac{1\pm i \sqrt{3}}{2}
  - \sqrt[3]{\zeta } \frac{1\mp i \sqrt{3}}{2}  +\Omega  (2
  q-9),
  \label{N}
\end{equation}
and  $\alpha_{n} = \Omega+in\sqrt{2J^2-\Omega^2}$ and $\zeta =
\sqrt{\nu ^2+\phi ^3}+\nu,$ together with $\phi = 54 J^2+\Omega^2
\left(-4 q^2+18 q-27\right)$ and $\nu = \Omega q \left[324
J^2+\Omega^2 \left(8 q^2-54 q+81\right)\right]$. A key observation
is that the first six eigenvalues remain independent of the
parameter $q$, demonstrating an exact correspondence between the
NHH and the full Liouvillian. Furthermore, for $q=0$, the
expression simplifies to $\zeta = \left(54 J^2-27
\Omega^2\right)^{3/2}$. This leads to the following spectrum of
the NHH:
\begin{equation}
\left\{0,-2 \Omega
,-\alpha_1^{*},-\alpha_1^{*},-\alpha_{1},-\alpha_{1},
-2\alpha_{1}, -2\alpha_{1}^{*},-2\Omega \right\},
\label{eq:NHH_spectrum}
\end{equation}
which reveals the emergence of a third-order exceptional point
(EP3) in the NHH (see Fig.~\ref{fig:tot_spec_k1}), marked by the
coalescence of the three highest eigenvalues and their associated
eigenvectors at $J \approx 21.1$. At first glance, it may appear
counterintuitive that transitioning to the superoperator formalism
enhances the order of the EP from second to third. Although this
behavior is not universal, it is well-justified. The superoperator
representation effectively enlarges the dimensionality of the
system's state space, which can lead to higher-order spectral
degeneracies. A detailed discussion of the relationship between
the spectra of non-Hermitian Hamiltonians and their corresponding
superoperators is provided in Appendix~\ref{app:D_correspondance}.

\subsection{The spectral role of quantum jumps}
\label{sec:no_RF_case_jumps}

\begin{figure*}[ht]
    \centering
    \includegraphics[width=0.9\textwidth]{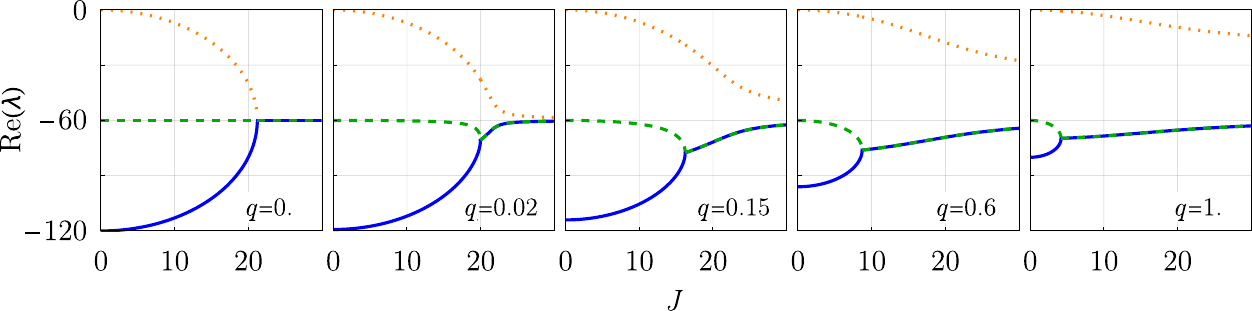}
\caption{Real part of the hybrid Liouvillian spectra as a function
of $J$, calculated from Eq.~\eqref{eq:hybrid_Liouvillian_matrix}
for various values of the quantum-jump parameter $q$. The
displayed eigenvalues correspond to those that form EP3 in the NHH
limit. The plots illustrate how increasing $q$ gradually lifts the
spectral degeneracy, thereby modifying the spectrum. Different
panels show results for $q$ ranging from 0 (purely non-Hermitian
case) to 1 (full Liouvillian case). We set $\Omega = 30$.}
    \label{fig:re_spec_full_parts}
\end{figure*}

Analysis of the results shown in Fig.~\ref{fig:re_spec_full_parts} suggests that the inclusion of quantum jumps effectively lifts degeneracies, reducing the third-order HEP to a second-order LEP. To confirm this, one can calculate the quasienergy splitting of the eigenvalues as
\begin{equation}
    \Delta E_{ij}^{R} = \text{Re}\left(\lambda'_i - \lambda'_j\right), \qquad
    \Delta E_{ij}^{I} = \text{Im}\left(\lambda'_i - \lambda'_j\right).
\end{equation}
Since the spectra of both the Liouvillian and hybrid Liouvillian are generally complex--reflecting the non-Hermitian nature of these superoperators--it is essential to analyze the real and imaginary parts of the quasienergy splitting separately.

\begin{figure}[ht]
   \includegraphics[width=\columnwidth]{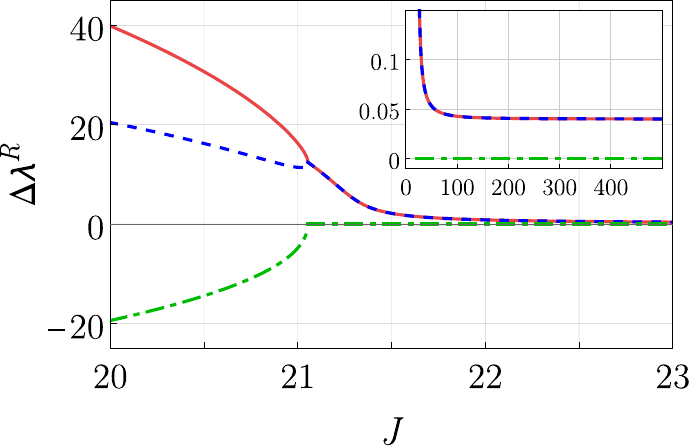}
    \caption{
Real and imaginary parts of the quasienergy splittings between the
eigenvalues that originally formed the EP3 at $q=0$, calculated
from Eq.~\eqref{eq:hybrid_Liouvillian_matrix}. The splittings
shown are $\Delta \lambda^{R}_{89}$ (green dot-dashed curve),
$\Delta \lambda^{R}_{79}$ (red solid curve), and $\Delta
\lambda^{R}_{78}$ (blue dashed curve). The chosen value $q =
0.001$ demonstrates how even a small quantum jump contribution
lifts the degeneracy and reduces the EP order. The inset displays
the real part of the splitting over a wider range of $J$,
highlighting the asymptotic behavior. We set $\Omega = 30$.}
    \label{fig:energy_splitting}
\end{figure}

Particularly interesting is the analysis of the three eigenvalues
$\lambda_{7,8,9}$ that form the EP3. From Fig.~\ref{fig:energy_splitting}, it is evident that
$\Delta E^{R(I)}_{89}$ reaches zero sharply around $J \approx
21.1$. In contrast, the behavior of $\Delta E^{R(I)}_{78}$ and
$\Delta E^{R(I)}_{79}$ is less clear, as these appear to approach
zero asymptotically. To confirm this trend, we evaluate the limit
of $\Delta E^{R}_{ij}$ as $J \to +\infty$:
\begin{eqnarray}
    \lim_{J\rightarrow +\infty}\Delta E^{R}_{78} &=& \lim_{J\rightarrow +\infty}\Delta E^{R}_{79} = \dfrac{4}{3}\Omega q,\\
    \lim_{J\rightarrow +\infty}\Delta E^{R}_{89} &=& 0.
\end{eqnarray}
This result confirms that even a small contribution of quantum
jumps lifts the degeneracy of the third-order HEP. For the
parameters used in Fig.~\ref{fig:energy_splitting}, the asymptotic
splitting is approximately $0.04$.

\subsection{Interpretation of the quantum-jump parameter}

\begin{figure}[ht]
    \centering   \includegraphics[width=\linewidth]{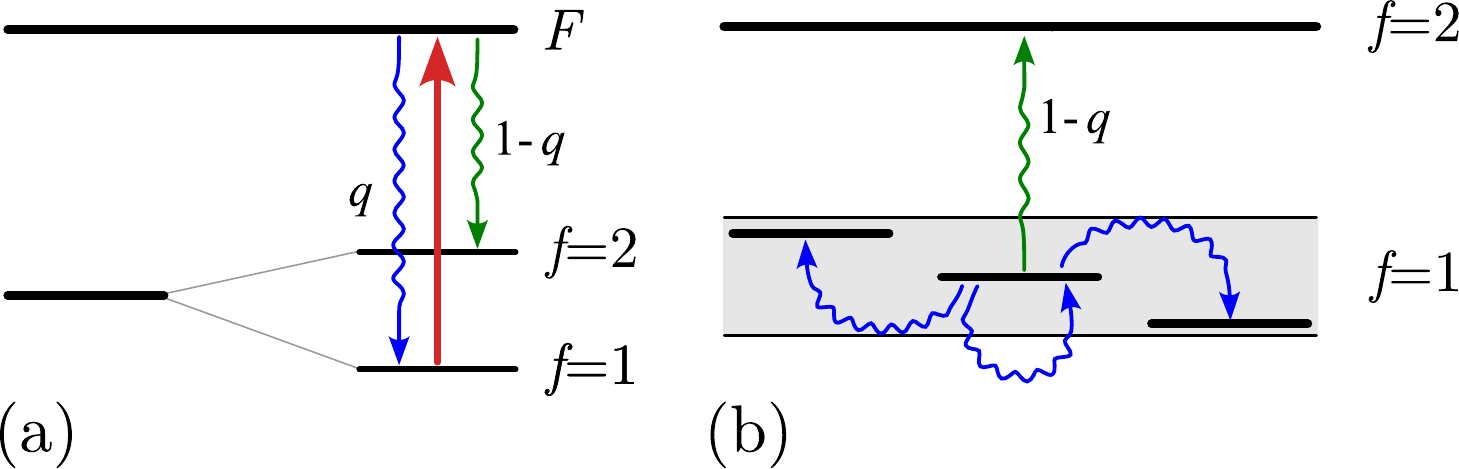}
    \caption{
Schematic illustration of the physical interpretation of the
parameter $q$ in the dynamics of atomic systems. (a) The full
atomic level structure, where $q$ denotes the fraction of the
excited-state population that decays into the monitored manifold
($f = 1$), while the remaining fraction $(1 - q)$ decays into an
unobserved manifold ($f = 2$). (b) An effective model in which the
monitored manifold ($f = 1$) is treated as a reduced system. Here,
$(1 - q)$ characterizes an effective population loss rate into the
unobserved manifold ($f = 2$).}
    \label{fig:hybrid_Liouvillian}
\end{figure}

While the quantum-jump parameter $q$ serves as a valuable
mathematical tool that enables a smooth interpolation between the
spectra of an NHH and the full Liouvillian superoperator, it also
possesses a clear physical interpretation \cite{Minganti2020,
Chen2021, Kumar2021}. In the context of atomic vapors, this
interpretation becomes evident from
Fig.~\ref{fig:hybrid_Liouvillian} and the structure of the master
equation~\eqref{eq:03_master_equation}. As detailed in
Appendix~\ref{app:A4}, in atomic, molecular, and optical (AMO)
physics, the final term in the master equation---associated with
quantum jumps---is commonly referred to as the repopulation term,
which ensures the preservation of the normalization of the reduced
system's density matrix during open-system evolution.

From this perspective, the factor $(1 - q)$ quantifies the leakage
of probability current into unobserved subspaces of the system. In
the example shown in Fig.~\ref{fig:hybrid_Liouvillian}, this
corresponds to population decay into the second spin manifold ($f
= 2$), which remains unmonitored during the measurement process.

An analogous interpretation of $q$ arises in quantum circuit
dynamics experiments, where non-Hermitian evolution is typically
engineered through post-selection on quantum trajectories. In such
settings, the parameter $q$ effectively captures the detection
efficiency---i.e., the likelihood of successfully monitoring
quantum jumps. A value of $q < 1$ thus reflects imperfect
detection or deliberate disregard of transitions into certain
states considered as unobserved.

\section{RF Detuning Regime of the Model}
\label{sec:RF_case}

We consider a natural extension of the model of
Ref.~\cite{Liang2023} for the $f=1 \rightarrow F=0$ transition,
now assuming that the RF field is slightly detuned from exact
resonance with the Larmor frequency. In any case, we continue to
assume zero optical detuning. Since the physical system remains
the same as described in Sec.~\ref{sec:model_eff_Hamiltonian}, we
can directly apply Eq.~\eqref{eq:03_eff_ham_article} with $\Delta
= 0$. Thus we have
\begin{equation}
    \NHH = \matrix{\delta&J&0 \\ J&-2i \Omega&J \\ 0&J&-\delta },
    \label{eq:03_Hnh_detuned}
\end{equation}
where $\delta$ denotes the detuning of the RF field from the Larmor frequency, introducing an additional degree of freedom for controlling and manipulating the system.

\subsection{Spectrum of the Hamiltonian operator}
\label{sec:RF_case_operator}

\begin{figure}
    \centering
    \includegraphics[width=\columnwidth]{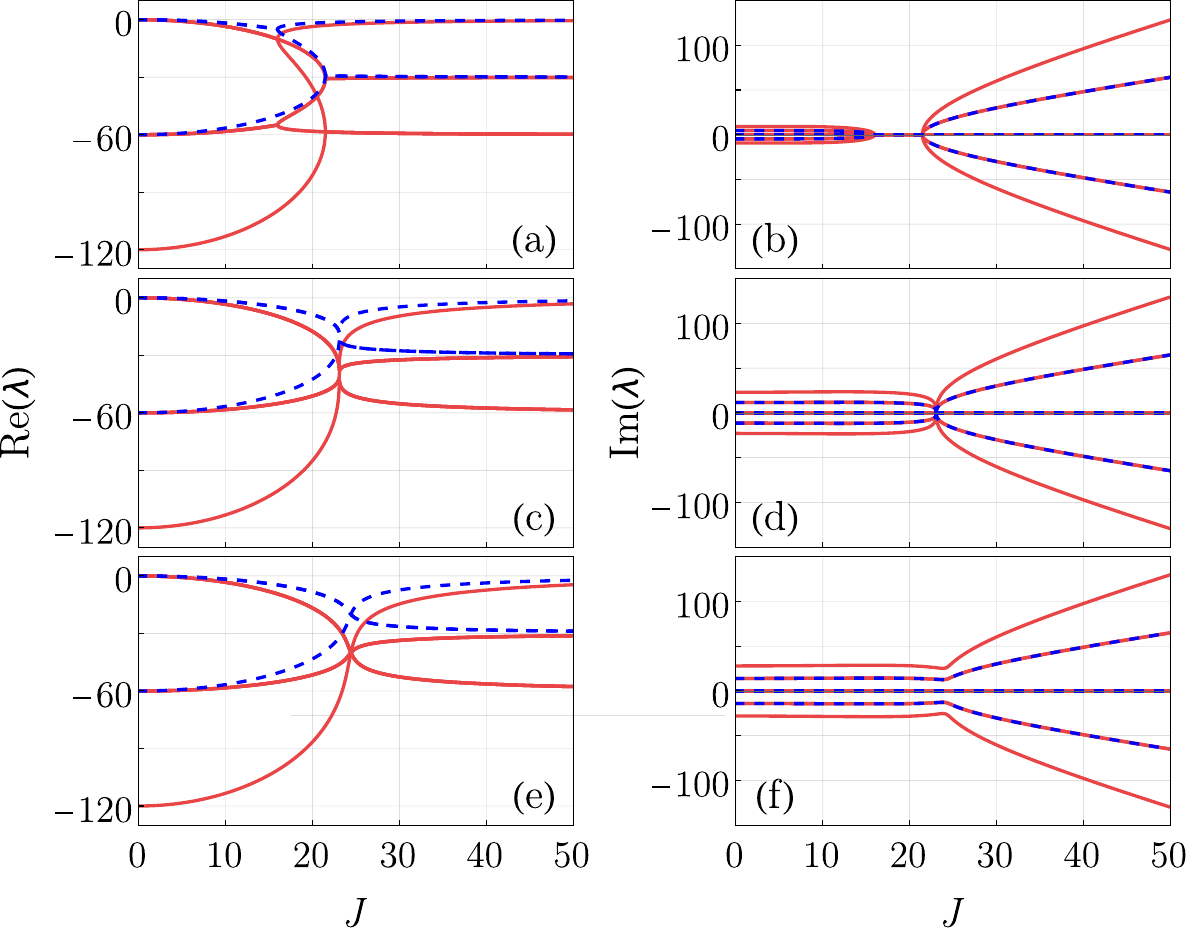}
\caption{Hamiltonian spectra of the three-level system as a
function of the RF field detuning $\delta$ from the Larmor
frequency $\Omega_L$. (a, b): $|\delta| < \delta_0\equiv 2\Omega /
(3\sqrt{3})$; (c, d): $|\delta| > \delta_0$; (e, f): At the
critical point $\delta = \delta_0$. Red curves show the spectra of
the NHH superoperator [obtained from
Eq.~\eqref{eq:hybrid_Liouvillian_matrix} in the limit $q=0$],
while blue dashed curves correspond to the spectra of the
associated operator [calculated from
Eq.~\eqref{eq:03_Hnh_detuned}]. The left column presents the real
part of the spectra, and the right column shows the imaginary
part. Assumed parameters: $\Omega = 30$, with $\delta \approx
4.62$, $11.55$, and $14$, respectively.}
    \label{fig:detuned_3_regions}
\end{figure}

A straightforward spectral analysis shows that the eigenvalues of
the NHH are given by the roots of the characteristic polynomial:
\begin{equation}
    x^3 + 2i \Omega x^2 - (2J^2+\delta^2) x-2i\Omega \delta = 0,
    \label{eq:03_characteristic_eq}
\end{equation}
Analytical solutions to this cubic equation are generally not straightforward. However, with three adjustable parameters ($\Omega$, $J$, and $\delta$), one can tailor the system to exhibit specific spectral features. Notably, it is possible to realize a third-order degeneracy, where all three eigenvalues coalesce at a single point. To identify such a point, we assume the characteristic polynomial factorizes as $(x - i x_0)^3 = 0$. Given that the constant and quadratic terms in Eq.~\eqref{eq:03_characteristic_eq} are purely imaginary, the eigenvalues at this degeneracy must also be purely imaginary. This assumption leads to the following set of conditions:
\begin{eqnarray}
    x_0 &=& -\frac{2}{3} \Omega,\\
    x_0^2 &=& \frac{1}{3}\left( 2J^2 + \delta^2 \right),\\
    x_0^3 &=& -2 \Omega \delta^2.
\end{eqnarray}
Solving this system of equations yields four solutions sharing the
same quasienergy, given by $E_{\rm{tp}} = i x_0 = -2 i \Omega/3$.
Assuming $J > 0$ (since $J$ represents the RF Rabi frequency), we
also obtain two symmetric solutions for the detuning parameter:
\begin{equation}
    \delta = \pm \dfrac{2}{3 \sqrt{3}}\Omega, \qquad J=\dfrac{4}{3\sqrt{3}} \Omega.
    \label{eq:03_triple_point}
\end{equation}
This symmetry reflects the system's invariance under reversal of
the magnetic field, which corresponds to changing the sign of the
detuning or, equivalently, exchanging the states $\ket{0}
\leftrightarrow \ket{2}$ [see Fig.~\ref{fig:energy_level_diagram}
and Eqs.~\eqref{eq:03_time_dep_hamiltonian} and
\eqref{eq:03_time_indep_hamiltonian}].

Furthermore, it can be shown that within the detuning range
$\abs{\delta} < (2/3\sqrt{3}) \Omega$, there exist two
distinct values of $J$ at which second-order degeneracy occur,
corresponding to a HEP, as discussed in detail in the next
section. Outside this interval, no degeneracies are observed. This
behavior demonstrates that by varying the RF field detuning, it is
possible to completely transform the system's
characteristics---from having two second-order EPs (EP2), through
a single third-order EP (EP3), to a regime where no exceptional
points are observed (see Fig.~\ref{fig:detuned_3_regions}).

An analysis of the eigenvector orthogonality confirms that all observed degeneracies correspond to EPs. In particular, at a EP3, all eigenstates coalesce into a single, parameter-independent eigenvector:
\begin{equation}
    \ket{E_{\rm{tp}}} = \frac{1}{6} \left(\sqrt{3}+3 i\right)\ket{1\bar{1}}+\frac{1}{6} \left(\sqrt{3}-3 i\right)\ket{10}+\frac{1}{\sqrt{3}}\ket{11}.
\end{equation}

\begin{figure}[t]
    \centering
     \includegraphics[width=\columnwidth]{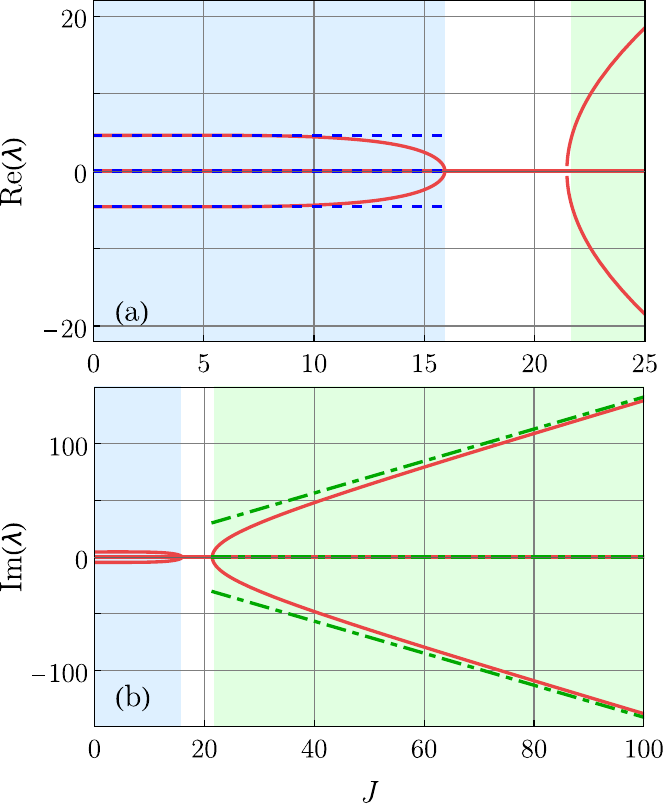}
    \caption{
(a) Real and (b) imaginary parts of the spectrum of the NHH
operator (red solid curves), calculated from
Eq.~\eqref{eq:hybrid_Liouvillian_matrix}, illustrating the asymptotic behavior. The
blue-shaded region ($J \to 0$) corresponds to the dynamics
dominated by the $\delta \hat{F}_z$ term where $\hat{F}_i$ is the
angular momentum operator along the $i$ axis, with the associated
asymptote shown as a blue dashed line in panel (a). The
green-shaded region ($J \to \infty$) reflects the dynamics
governed by the $J \hat{F}_x$ term, with the corresponding
asymptotes shown by green dot-dashed lines in panel (b). We set
$\Omega = 30$ and $\delta = 4.62$.}
    \label{fig:asymptotical}
\end{figure}

\subsection{Physical interpretation of the spectral regions}

To elucidate the physical significance of these three regions, we
examine the asymptotic behavior of the Hamiltonian in
Eq.~\eqref{eq:03_Hnh_detuned} and its influence on the real part
of the spectrum. The regime $|\delta| < 2\Omega/(3\sqrt{3})$ is
particularly revealing, as it features three distinct intervals in
$J$, each corresponding to a different dominant dynamical
behavior.

As shown in Fig.~\ref{fig:asymptotical}(a), for small values of
$J$, the spectrum approaches that of the Hamiltonian
$\hat{H}_{J\to 0} = \delta \hat{F}z$, where $\hat{F}_i$ is the
angular momentum operator along the $i$ axis. This indicates that
the detuning term is dominant. In this regime, optical pumping
(characterized by $\Omega$) primarily influences the imaginary
part of the spectrum, while the real part remains nearly
unaffected. In the opposite limit ($J \to \infty$), the spectrum
converges to that of $\hat{H}_{J \to \infty} = J \hat{F}_x$,
signifying that the RF field governs the system's dynamics.

The intermediate regime is more intricate. Previous analysis at
$\delta = 0$ shows that a bifurcation emerges when $J \ge \Omega /
\sqrt{2}$, marking the transition from optical-pumping-dominated
behavior to RF-field dominance. Furthermore, the coupling
parameter $J$ links the dark states ($\ket{11}$ and
$\ket{1\bar{1}}$) to the optically active state, thereby enhancing
the influence of light and enabling it to affect not only the
imaginary but also the real part of the spectrum. Finally, in the
regime where the light-induced non-Hermitian component dominates,
the eigenvalues are expected to be predominantly imaginary.

In summary, we conjecture that the three observed regions
correspond to: (1) a detuning-dominated regime, where $\delta$
exceeds the influence of both $J$ and $\Omega$; (2) a
light-dominated regime, where $J$ is small enough for optical
effects to dominate over $\delta$, yet not strong enough to
suppress them; and (3) an RF-dominated regime, where $J$ becomes
sufficiently large to overpower both optical and detuning effects.

This framework also explains the absence of EPs in
Fig.~\ref{fig:detuned_3_regions}(c): the optical field is too
weak relative to $\delta$, and while a large $J$ is needed to
induce eigenvalue coalescence, such a value simultaneously
suppresses all competing effects---including the optical ones.

\begin{figure}[t]
    \centering
     \includegraphics[width=\columnwidth]{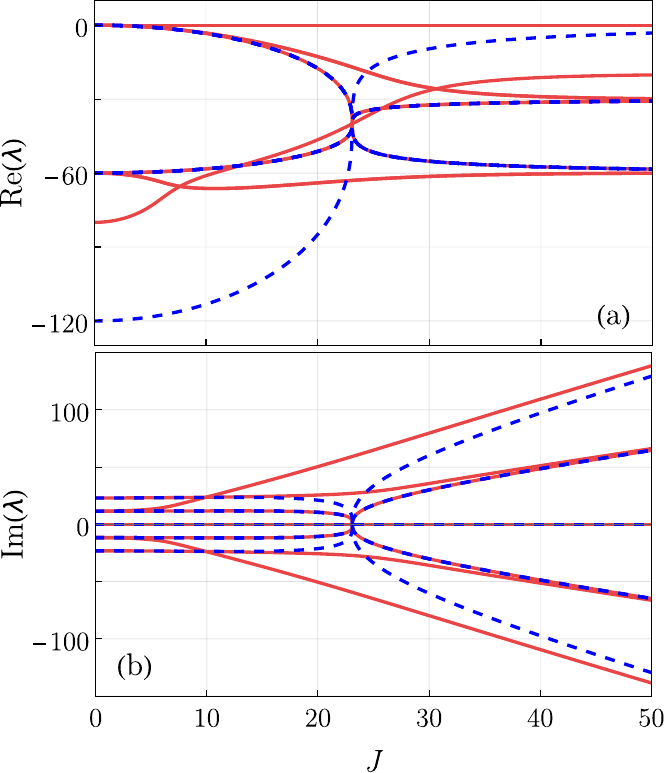}
\caption{(a) Real and (b) imaginary parts of the spectra of the
Liouvillian (red solid curves) and the NHH (blue dashed curves) as
a function of the coupling strength $J$, shown for parameters
corresponding to the emergence of the EP3. The spectra correspond
to the limits $q = 1$ and $q = 0$ of
Eq.~\eqref{eq:hybrid_Liouvillian_matrix_2}, respectively. This figure illustrates how the inclusion of quantum jumps
modifies the spectrum and lifts the eigenvalue degeneracies. We
set: $\Omega = 30$ and $\delta \approx 11.55$. }
    \label{fig:liouv_hnh_triple}
\end{figure}

\subsection{Superoperator representation}
\label{sec:RF_case_superoperator}

The results presented here constitute a natural generalization of
those in Sec.~\ref{sec:no_RF_case}. As before, to facilitate a
straightforward comparison between the full Liouvillian spectrum
and that of the non-Hermitian Hamiltonian, we adopt the same
hybrid Liouvillian technique. Specifically, the hybrid Liouvillian
introduced in Eq.~\eqref{eq:hybrid_Liouvillian_matrix} generalizes
to:
\begin{widetext}
\begin{equation}
    \dhat{\mathcal{L}}(q) = \left(
\begin{array}{ccccccccc}
 -2 \Omega  & -\delta  & 0 & 0 & J & 0 & 0 & 0 & 0 \\
 \delta  & -2 \Omega  & -2 J & -J & 0 & 0 & 0 & 0 & 0 \\
 0 & 2 J & -2 \Omega  & 0 & 0 & 0 & -J & \frac{2 \Omega }{\sqrt{3}} & 2 \sqrt{\frac{2}{3}} \Omega  \\
 0 & J & 0 & 0 & -2 \delta  & 0 & -J & 0 & 0 \\
 -J & 0 & 0 & 2 \delta  & 0 & J & 0 & 0 & 0 \\
 0 & 0 & 0 & 0 & -J & -2 \Omega  & -\delta  & 0 & 0 \\
 0 & 0 & J & J & 0 & \delta  & -2 \Omega  & -\sqrt{3} J & 0 \\
 0 & 0 & \frac{2 \Omega }{\sqrt{3}} & 0 & 0 & 0 & \sqrt{3} J & -\frac{2}{3} \Omega & -\frac{2}{3} \sqrt{2} \Omega  \\
 0 & 0 & 2 \sqrt{\frac{2}{3}} \Omega (q-1)  & 0 & 0 & 0 & 0 & -\frac{2}{3} \sqrt{2} \Omega (q-1)  & -\frac{4}{3} \Omega (q-1) \\
\end{array}
\right).
\label{eq:hybrid_Liouvillian_matrix_2}
\end{equation}
\end{widetext}
where a similar pattern of changes can be observed as in the
previous case, particularly in the modification of the last row,
which ensures the conservation of state normalization.

As in the case of the standard form of the detuned Hamiltonian,
the general expressions for the quasienergies and eigenvectors are
rather complicated and do not offer significant physical insight.
However, as shown in previous section the standard form of the NHH
reveals the existence of a particularly notable singularity within the
three-dimensional parameter space $\Omega$, $\delta$, and $J$ [see
Eq.~\eqref{eq:03_triple_point}], in which the EP3 arises. This
point serves as a natural candidate for probing discrepancies
between the NHH and the full Liouvillian descriptions, as the NHH
superoperator is expected to inherit certain features from the
standard-form dynamics. For these parameter values, the
corresponding spectra are shown in
Fig.~\ref{fig:liouv_hnh_triple}. It can be observed that within
the NHH superoperator description, the entire NHH spectrum
collapses, with all nine eigenvalues becoming degenerate.
Furthermore, the analysis of the Jordan chain reveals the presence
of only three linearly independent eigenvectors, indicating the
existence of at least two HEPs of orders 3 and 5. Introducing
quantum jumps reduces the degeneracy order at this point
significantly, demonstrating a marked decrease in the degree of
degeneracy.

\section{Proposal for experimental observation of
Liouvillian exceptional points}

In this subsection, we briefly outline possible methods for the
experimental observation of the key results of our work---namely,
the LEPs and their generalized forms, $q$-dependent LEPs, in the
studied system.

Since LEPs correspond to singularities in the Liouvillian
spectrum, the most natural and complete method for detecting them
is to perform full quantum process tomography (QPT), which enables
reconstruction of the entire Liouvillian superoperator. For
systems with a few qubits implemented in superconducting quantum
circuits, such a QPT-based approach was recently employed on the
IBMQ platform to reveal single-qubit LEPs \cite{Abo2024}.

\begin{figure}
    \centering
    \includegraphics[width=\linewidth]{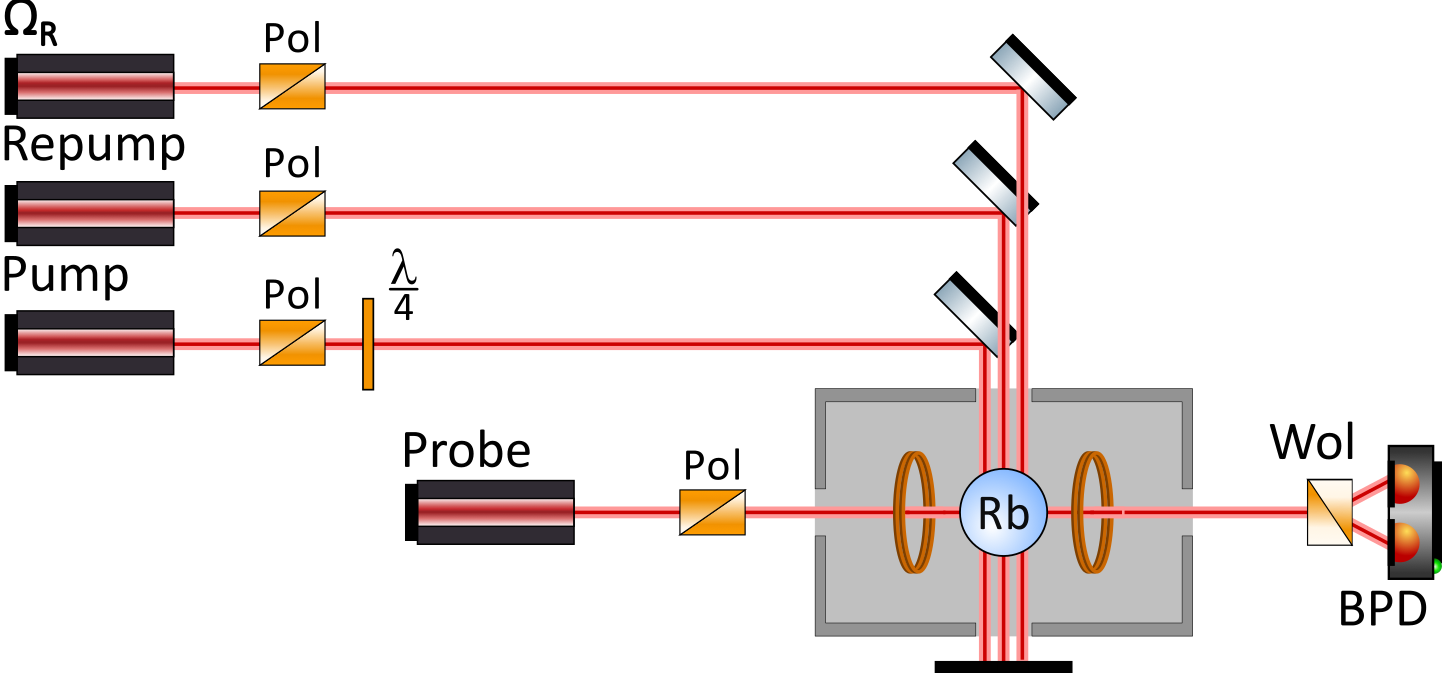}
    \caption{
Conceptual schematic of the experimental setup used for quantum
process tomography (QPT) in room-temperature atomic vapors,
enabling the observation of LEPs and hybrid LEPs. Three laser
beams---labeled Pump, Repump, and Probe---are used to implement
the tomography protocol described in Refs. \cite{Kopciuch2024,
Sun2025}, which relies on free-induction decay (FID)-based state
reconstruction \cite{Kopciuch2024}. An additional laser beam,
denoted $\Omega_R$, provides the optical coupling required by the
theoretical model. Optical components include: Pol -- a
crystalline polarizer; $\lambda/4$ -- a quarter-wave plate; Wol --
a Wollaston prism; and BPD -- a balanced photodetector. The Rb
cell refers to a paraffin-coated glass vapor cell containing
rubidium-87, enclosed within a magnetically shielded environment.
}
    \label{fig:experimental_setup}
\end{figure}

In our case, however, we consider atomic vapor systems and a
qutrit (rather than a qubit). While the general concept remains
the same, we propose to adopt a QPT method adapted to atomic
vapors, based on quantum state tomography (QST) as introduced in
\cite{Kopciuch2022} and experimentally implemented in
\cite{Kopciuch2024}. Figure~\ref{fig:experimental_setup} shows a
simplified scheme of the experimental setup used for QST in
\cite{Kopciuch2024}.

To extend this QST protocol to full QPT, it is sufficient to apply
QST to a complete set of input basis states. The experimental
setup can remain unchanged from that in
Fig.~\ref{fig:experimental_setup}. For example, in the case of a
qutrit, the basis states can be taken as eigenstates of the eight
Gell-Mann matrices, which span the SU(3) algebra~\cite{Thew2002}.
By performing QST on the output states corresponding to each of
these basis inputs, full reconstruction of the Liouvillian becomes
possible. One can then extract its eigenvalues and search for
degeneracies using the same strategy as in \cite{Abo2024}.

Regarding the observation of hybrid Liouvillians, we note that
several variants of QPT exist, including so-called Lindblad
tomography, which enables separate reconstruction of the coherent
(Hamiltonian) and incoherent (dissipator) parts of the dynamics.
This method was demonstrated experimentally for reconstructing
single- and two-qubit Lindbladians on a superconducting quantum
processor in \cite{Samach2022}. Through appropriate
post-processing of QPT data, it is possible to isolate the
Lindblad dissipator and Hamiltonian contributions. By adjusting
the statistical weight parameter $q$ between them, hybrid
Liouvillians can be constructed and analyzed.

An alternative, more direct route to observing system dynamics for
a specific value of $q$ is to engineer post-selected trajectories.
In optical systems, this is often simulated by adjusting detection
efficiency using beam splitters or attenuators \cite{Minganti2020,
Kumar2021}. Another method is to monitor a specific dissipative channel via an
ancillary detector, and post-select only those measurement runs
with a specified number of detected quantum jumps \cite{Chen2021}.

In the case of atomic vapors, where the measured signal contains
ensemble-averaged information from many atoms, these jumps
manifest as a reduction in the free-induction decay (FID) signal,
including observable changes in absorption or polarization
rotation. This becomes more intuitive upon realizing that the
amplitude of the FID signal is proportional to the number of atoms
actively interacting with the light field~\cite{Kopciuch2022}.
Referring to Fig.~\ref{fig:hybrid_Liouvillian}, one can observe
that jumps into the unobserved subspace reduce the number of atoms
contributing to the measurable signal, thereby decreasing the FID
amplitude in a detectable manner.

We propose applying a similar approach to our atomic vapor system
by embedding the qutrit dynamics within a four-level (quartit)
system. As illustrated in Fig.~\ref{fig:hybrid_Liouvillian}(b),
the three lower levels---magnetic sublevels of the $f=1$ hyperfine
state---encode the qutrit states under investigation, while the
fourth level serves as an effective decay channel.

To simulate the hybrid Liouvillian with $q=0$ (corresponding to
purely non-Hermitian dynamics), the experiment must be designed
such that the excited state, to which the qutrit is optically
coupled, cannot decay into the unobserved part of the system. A
relevant example is the $f=1 \rightarrow F=0$ transition, where
subsequent spontaneous decay from $F=0$ to $f=2$ is forbidden by
selection rules. Intermediate values of $q$ can be obtained by
tuning the system to a different excited state, effectively
modifying the relative transition strengths between the qutrit
subspace ($f=1$) and the auxiliary manifold ($f=2$).

This approach enables the implementation of a conditional QPT
protocol tailored for reconstructing hybrid Liouvillians. The
experimental setup illustrated in
Fig.~\ref{fig:experimental_setup} allows QPT to be performed
across a tunable range of $q$ values, with the atomic dynamics
confined to the appropriate regime [see
Fig.~\ref{fig:energy_level_diagram}(b)]. In doing so, the
methodology developed in~\cite{Chen2021} for superconducting
quantum circuits is effectively adapted to room-temperature atomic
vapor systems, paving the way for experimental exploration of both
standard and hybrid LEPs in a qutrit.

\section{Conclusions}
\label{sec:conclusions}

In this work, motivated by the theoretical and experimental
results of Ref.~\cite{Liang2023}, we have explored spectral
singularities of their alkali-metal atomic vapor system (as shown
in Fig.~\ref{fig:energy_level_diagram}), modeled using four and
effectively three hyperfine states, to investigate the nature of
EPs in open quantum dynamics. By systematically comparing the
spectra of NHHs and quantum Liouvillian superoperators, we
identified significant discrepancies between semiclassical and
fully quantum descriptions of dissipation especially in the
detuned system.

Our results demonstrate that while NHHs can approximate the
system's behavior in limited regimes---particularly in the absence
of quantum fluctuations---their predictive power breaks down when
quantum jump processes become significant. These jumps, intrinsic
to the Lindblad framework, are responsible for state repopulation
and fundamentally reshape the system's spectral features. This is
particularly important for atomic systems with particle number
conservation, where repopulation terms are unavoidable for an
accurate description of the system. These terms give rise to many
important phenomena, such as optical pumping \cite{AuzinshBook}.

We have presented concrete examples in which the presence or
absence of EPs, their precise location in parameter space, or
their algebraic multiplicity differ markedly between the NHH and
Liouvillian approaches. These findings highlight that quantum
jumps do not merely perturb the spectrum but may induce or destroy
the order of degeneracies, altering the qualitative structure of
the dynamics.

To bridge the gap between these two approaches, we employed the
hybrid-Liouvillian formalism, which enables a controlled
interpolation between the jump-free and fully stochastic regimes.
This framework clarifies how spectral features predicted by NHHs
evolve under the inclusion of quantum jumps, ultimately converging
to the Liouvillian spectrum that governs fully quantum dynamics.

Regarding the experimental verification of the predicted
Liouvillian EPs, this can be achieved indirectly by generalizing
the approach of Ref.~\cite{Liang2023}, or directly and
comprehensively through quantum process tomography. The latter
approach follows the method recently demonstrated in a circuit QED
experiment~\cite{Abo2024}. In particular, the quantum
state~\cite{Kopciuch2022, Kopciuch2024} and process~\cite{Sun2025}
tomography techniques developed for alkali-metal atomic vapors
could, in principle, be adapted to the system studied in
Ref.~\cite{Liang2023}.

Overall, our study reinforces the importance of Liouvillian-based
methods for accurately capturing the spectral singularities of
open quantum systems, particularly in effectively low-dimensional
atomic platforms where quantum noise cannot be neglected. These
insights are relevant for ongoing experimental efforts in quantum
optics, quantum thermodynamics, and quantum sensing, where
Liouvillian EPs offer both fundamental and practical significance.

\appendix

\section{Effective system description}
\label{app:A_effective_system_description}

The calculation of effective dynamics can be challenging in
certain cases. Here, we apply the method introduced in
Ref.~\cite{Reiter2012}. This formalism is valid under several key
assumptions: (1) the system must exhibit Markovian dynamics, (2)
be described within a Liouvillian framework, (3) feature a
perturbative coupling between ground and excited states, and (4)
display a clear separation of timescales---specifically, the
relaxation of the excited state must occur much faster than the
evolution within the ground-state manifold.

While the original formulation does not explicitly include the
intrinsic, slow relaxation of the ground state, it can be
naturally extended to incorporate this effect. In our system, we
identify two distinct types of Lindblad operators. The first,
$\hat{L}_{\varepsilon}^{\rm{sp}}$, represents spontaneous emission from the
excited state with polarization $\varepsilon \in {\pm1, 0}$. The second,
$\hat{L}_{\mu}^g$, accounts for the intrinsic slow relaxation
processes within the ground-state manifold. Importantly, since
$\hat{L}_{\mu}^g$ does not couple to the excited state, it remains
unaffected by the effective operator formalism. In contrast,
$\hat{L}_{\varepsilon}^{\rm{sp}}$ must be modified to reflect the effective
dissipation in the ground state, arising from transitions
through the excited state followed by spontaneous emission.

We begin by decomposing the system Hamiltonian as:
\begin{equation}
    \hat{H} = \hat{H}_{g} + \hat{H}_{e} + \hat{V}_+ + \hat{V}_-,
\end{equation}
where $\hat{H}_{g(e)} = \hat{P}_{g(e)} \hat{H} \hat{P}_{g(e)}$
represents the ground (excited) state Hamiltonian, obtained via
the projection operators $\hat{P}$. Additionally, $\hat{V}_{+(-)}
= \hat{P}_{e(g)}\hat{H}\hat{P}_{g(e)}$ corresponds to the
generalized raising and lowering operators, respectively.

To formalize this approach, we introduce the effective
non-Hermitian Hamiltonian governing the dynamics within the
excited-state manifold as:
\begin{equation}
    \hat{H}_{\rm{eNH}} = \hat{H}_{e} - \frac{i}{2}\sum_{\varepsilon}(\hat{L}_{\varepsilon}^{\rm{sp}})^{\dagger}\hat{L}_{\varepsilon}^{\rm{sp}},
    \label{eq:03_excited_nh_ham}
\end{equation}
where, following Ref.~\cite{AuzinshBook}, the Lindblad operators
are proportional to the electric-dipole-moment operator governing
transitions between the ground and excited states, while also
accounting for the directionality of the emission, since
spontaneous emission cannot excite the atom---i.e.,
$\hat{L}_{\varepsilon}^{\rm sp} \propto
\hat{P}_g\hat{d}_\varepsilon\hat{P}_e$. With this in mind, one can
define the effective ground-state Hamiltonian and the effective
spontaneous emission Lindblad operators as
\begin{eqnarray}
    \label{eq:03_eff_ham}
    \hat{H}_{\rm{eff}} &=& -\dfrac{1}{2}\hat{V}_{-} \left[ H_{\rm{eNH}}^{-1} + \left(\hat{H}_{\rm{eNH}}^{-1} \right)^{\dagger} \right] \hat{V}_{+} + \hat{H}_{g},\\
    \label{eq:03_eff_Lindblad}
    \hat{L}_{\rm{eff}}^{\varepsilon} &=& \hat{L}_{\varepsilon}^{\rm{sp}}\hat{H}_{\rm{eNH}}^{-1}\hat{V}_{+}.
\end{eqnarray}
It is worth noting that, in contrast to other methods, such an
effective Hamiltonian does not require Hermitization to describe
unitary evolution, as it is Hermitian \emph{a~priori}, while all
effective dissipative processes are contained in
$\hat{L}_{\rm{eff}}^{\varepsilon}$ and $\hat{L}^{g}_{\mu}$, resulting in the
effective master equation or effective Liouvillian:
\begin{eqnarray}
    \dot{\rho}_g &=& \mathcal{L}_{\rm{eff}}(\rho_g) = -i \comm{\hat{H}_{\rm{eff}}}{\rho_g}
    \nonumber \\
    &&- \sum_{\varepsilon} \left( \dfrac{1}{2} \acomm{(\hat{L}^{\varepsilon}_{\rm{eff}})^{\dagger}\hat{L}^{\varepsilon}_{\rm{eff}} }{\rho_g} - \hat{L}^{\varepsilon}_{\rm{eff}}\rho_g(\hat{L}^{\varepsilon}_{\rm{eff}})^{\dagger}  \right)
    \nonumber \\ &&- \sum_{\mu} \left( \frac{1}{2} \acomm{(\hat{L}^{g}_{\mu})^{\dagger}\hat{L}^{g}_{\mu} }{\rho_g} - \hat{L}^{g}_{\mu}\rho_g(\hat{L}^{g}_{\mu})^{\dagger}  \right),
\end{eqnarray}
where $\rho_g$ denotes the density matrix projected onto the
Hilbert space of the ground-state manifold.

\subsection{Spontaneous emission} 
\label{app:A1}

As mentioned in the previous section, in the system of interest
the only relaxation channel that couples the excited and ground
states is spontaneous emission. Following Ref.~\cite{AuzinshBook},
the Lindblad operators describing this process are proportional to
the electric-dipole-moment operator governing transitions between
the ground and excited states. They must also account for
directionality, as spontaneous emission cannot excite the
atom---i.e., $\hat{L}_{\varepsilon}^{\rm{sp}} \propto
\hat{P}_g\hat{d}_{\varepsilon}\hat{P}_e$. In the case of a simple two-level
system, the total relaxation rate $\Gamma$ can be treated as an
experimentally determined parameter, reducing the problem to
finding the relative coupling strengths between different
sublevels:
\begin{equation}
\begin{split}
    \hat{L}^{\rm{sp}}_{\varepsilon} &= \sum_{m,M}\dfrac{\sqrt{\Gamma}}{\mel{f}{|\hat{d}|}{F}} \mel{fm}{\hat{d}_{\varepsilon}}{FM}
    \dyad{fm}{FM}\\
    &= \sqrt{\Gamma} \sum_{m,M} \matrix{f&1&F\\-m&\varepsilon&M} \dyad{fm}{FM},
\end{split}
\end{equation}
where we adopt the convention that lowercase (uppercase) letters
denote ground (excited) states. In the first equation, the
relaxation rate is rescaled by the reduced matrix element of the
dipole operator, $\mel{f}{|\hat{d}|}{F}$, which arises naturally
from the Wigner-Eckart theorem. This follows the convention in
Ref.~\cite{AuzinshBook} and is expressed using the Wigner 3j
symbol.

Since the reduced matrix element of $\hat{d}$ is not guaranteed to
be positive, particular care must be taken when evaluating the
Hermitian conjugate of the corresponding Lindblad operator:
\begin{equation}
\begin{split}
    (\hat{L}_{\varepsilon}^{\rm{sp}})^{\dagger} &= \sum_{m,M} \dfrac{\sqrt{\Gamma}}{\mel{f}{|\hat{d}|}{F}} \mel{FM}{\hat{d}_{\varepsilon}^{\dagger}}{fm}
    \dyad{FM}{fm}\\
    &=\sum_{m,M} \dfrac{\sqrt{\Gamma}}{\mel{f}{|\hat{d}|}{F}} \mel{FM}{\hat{d}_{-\varepsilon}}{fm}
    \dyad{FM}{fm}\\
    &=\dfrac{\sqrt{\Gamma}\mel{F}{|\hat{d}|}{f}}{\mel{f}{|\hat{d}|}{F}}  \sum_{m,M}
    \matrix{F&1&f\\-M&-\varepsilon&m} \dyad{FM}{fm}.
\end{split}
\end{equation}
By applying Eq.~(10.27) from Ref.~\cite{AuzinshBook} together with
the symmetry properties of the Wigner 3j
symbol~\cite{MessiahBook}, we obtain:
\begin{equation}
\begin{split}
    (\hat{L}_{\varepsilon}^{\rm{sp}})^{\dagger} &=\sqrt{\Gamma} (-1)^{F-f} \sum_{m,M}
    \matrix{F&1&f\\-M&-\varepsilon&m} \dyad{FM}{fm}\\
    &= \sqrt{\Gamma} (-1)^{2F+1} \sum_{m,M}
    \matrix{f&1&F\\m&-\varepsilon&-M} \dyad{FM}{fm}\\
    &= \sqrt{\Gamma} (-1)^{F-f} \sum_{m,M} \matrix{f&1&F\\-m&\varepsilon&M}
    \dyad{FM}{fm}.
\end{split}
\end{equation}
This result introduces a subtle ambiguity
because complex conjugation can alter the sign of a real-valued
operator. This issue arises since the phase factor is effectively
absorbed into the experimentally determined value of $\Gamma$.
Notably, due to selection rules for optical transitions, $\Delta f = F - f$ takes values
of $\pm1$ or $0$. To resolve this ambiguity, we redefine the jump
operators as follows:
\begin{equation}
    \label{eq:03_Lindblad}
    \hat{L}_{\varepsilon}^{\rm{sp}} = i^{\Delta f} \sqrt{\Gamma} \sum_{m,M}
    \matrix{f&1&F\\-m&\varepsilon&M} \dyad{fm}{FM}.
\end{equation}
This allows for a straightforward derivation of $(\hat{L}_{\varepsilon}^{\rm{sp}})^{\dagger}\hat{L}_{\varepsilon}^{\rm{sp}}$, which is useful, for example, in calculating the non-Hermitian Hamiltonian. Note that the above expression involves only the projection onto the excited state:
\begin{equation}
    (\hat{L}_{\varepsilon}^{\rm{sp}})^{\dagger}\hat{L}_{\varepsilon}^{\rm{sp}} = \Gamma \sum_{m,M}
    \matrix{f&1&F\\-m&\varepsilon&M}^2 \dyad{FM}{FM}.
    \label{eq:03_relax_q}
\end{equation}
In the calculation of the non-Hermitian Hamiltonians $\NHH$ or
$\hat{H}_{\rm{eNH}}$, we are especially interested in the
collective effect obtained by summing over all polarization
components $\varepsilon$:
\begin{equation}
\begin{split}
    \sum_{q} (\hat{L}_q^{\rm{sp}})^{\dagger}\hat{L}_q^{\rm{sp}} &= \Gamma \sum_{M}
    \left( \dyad{FM}{FM}
    \sum_{q,m}\matrix{f&1&F\\-m&q&M}^2 \right) \\
    &= \dfrac{\Gamma}{2F+1} \sum_{M} \dyad{FM}{FM}\\
    &= \dfrac{\Gamma}{2F+1} \hat{P}_{e},
    \label{eq:03_relax_tot}
\end{split}
\end{equation}
where we have used one of the summation rules for the Wigner 3j
symbol over magnetic sublevels~\cite{MessiahBook}.

\subsection{Raising and lowering operators}
\label{app:A2}

It is important to note that, in the model under consideration,
the primary mechanism coupling the ground and excited states is an
external light beam. Mathematically, this coupling closely
resembles the spontaneous-emission jump operator, as both
processes are governed by the same electric-dipole-moment
operator, $\hat{d}_\varepsilon$. Moreover, we treat the coupling light beam
as a classical field, which leads to the introduction of an
interaction term in the Hamiltonian, expressed as:
\begin{equation}
    \hat{H}_E = -\Vec{E} \cdot \hat{\Vec{d}} \cos(\omega t),
    \label{eq:03_light_interaction}
\end{equation}
where $\Vec{E} = E_0 \hat{e}$ represents the electric field vector
of the light beam, with amplitude $E_0$ and unit vector $\hat{e}$,
while $\omega$ is the carrier frequency.

Since the optical frequency is typically several orders of
magnitude higher than any other relevant frequency in the system,
the RWA naturally applies \cite{AuzinshBook}. For a simple
two-level system, the transformation between the static and
rotating frames is given by:
\begin{equation}
    \hat{U}_{\rm{rwa}} = \hat{P}_g + \hat{P}_e e^{-i \omega t}.
\end{equation}
Under this transformation, the Hamiltonian takes the form:
\begin{equation}
    \hat{\tilde{H}} = \hat{U}_{\rm{rwa}}^{\dagger} \hat{H} \hat{U}_{\rm{rwa}}
    - i\hbar \hat{U}_{\rm{rwa}}^{\dagger} \pdv{\hat{U}_{\rm{rwa}}}{t} = \hat{U}_{\rm{rwa}}^{\dagger} \hat{H} \hat{U}_{\rm{rwa}} - \hat{P}_e \hbar \omega.
    \label{eq:03_simple_rwa}
\end{equation}
This results in a simplified form of
Eq.~\eqref{eq:03_light_interaction}, which reads:
\begin{equation}
    \hat{H}_E = -\dfrac{1}{2}\Vec{E} \cdot \hat{\Vec{d}} - \hat{P}_e \hbar \omega = -\dfrac{E_0}{2} \sum_{\varepsilon} \hat{e}_{\varepsilon} \hat{d}_{\varepsilon} -\hat{P}_e \hbar \omega.
\end{equation}
For further details, see Ref.~\cite{AuzinshBook}. For simplicity,
we omit the tilde notation for transformed operators and refer to
the RWA only in the text.

To explicitly express this interaction, it is useful to apply the
Wigner-Eckart theorem. Moreover, since the dipole operator
$\hat{d}$ is an odd operator, it couples only ground and excited
states without inducing additional energy shifts. By applying
$\hat{V}_+ = \hat{P}_e \hat{H}_E \hat{P}_g$, we then obtain:
\begin{equation}
\begin{split}
    &\hat{V}_+ = -\dfrac{E_0}{2} \sum_{m,M} \mel{FM}{\sum_{\varepsilon} \hat{e}_{\varepsilon} \hat{d}_{\varepsilon}}{fm}
    \dyad{FM}{fm}\\
    &\;=-\dfrac{E_0}{2} \mel{F}{|\hat{d}|}{f} \sum_{\varepsilon,m,M} \hat{e}_\varepsilon \bar 1^{F-M}
    \matrix{F&1&f\\ \bar M&\varepsilon&m} \dyad{FM}{fm}\\
    &\;=\Omega_R \sum_{\varepsilon,m,M} \hat{e}_\varepsilon \bar 1^{F-M}
    \matrix{F&1&f\\ \bar M&\varepsilon&m} \dyad{FM}{fm},
\end{split}
\end{equation}
where $\Omega_R$ denotes the Rabi frequency, and for compactness,
we use the notation $\bar{1} = -1$ and $\bar{M} = -M$. This leads
to the total effect on the light beam being expressed as
\begin{eqnarray}
    \hat{V}_+ &=&  \Omega_R \sum_{\varepsilon,m,M} \hat{e}_q \bar 1^{\small F-M}
    \matrix{F&1&f\\ \bar M&\varepsilon&m} \dyad{FM}{fm},
    \nonumber \\
    \hat{H}_e &\rightarrow& \hat{H}_e -\hat{P}_e \hbar \omega.
\end{eqnarray}
and $\hat{V}_-=\hat{V}_+^{\dagger}$.

\subsection{Generalized rotating wave approximation}
\label{app:A3}

As mentioned in Sec.~\ref{sec:model_eff_Hamiltonian}, when the
system involves several distinct time-dependent interactions, it
is often more effective to apply a generalized form of the
rotating-wave approximation (RWA) rather than a sequence of
standard RWAs. This generalized transformation can be implemented
using the following unitary operation:
\begin{equation}
    \hat{U}_{\rm{grwa}} = \exp(-i \hat{G} t),
\end{equation}
where $\hat{G}$ is a time-independent generator of the
transformation. The choice of $\hat{G}$ is not unique; however, a
reasonable approach is to identify the frequency differences
between coupled states and construct $\hat{G}$ as a diagonal
operator encoding these detunings. Importantly, applying a
time-dependent basis transformation induces an effective energy
shift determined by $\hat{G}$:
\begin{equation}
\begin{split}
    \hat{\tilde{H}} &= \hat{U}_{\rm{grwa}}^{\dagger} \hat{H} \hat{U}_{\rm{grwa}} - i \hat{U}_{\rm{grwa}}^{\dagger}
    \pdv{\hat{U}_{\rm{grwa}}}{t}\\ &= \hat{U}_{\rm{grwa}}^{\dagger} \hat{H} \hat{U}_{\rm{grwa}} - \hat{G}.
\end{split}
\end{equation}
This implies that $\hat{G}$ should be chosen to simplify the
system's energy structure while minimizing the introduction of
unnecessary scalar terms in the Hamiltonian. In our case, the
simplest choice---yielding the desired form of the Hamiltonian, as
outlined in Ref.~\cite{Liang2023}---reads:
\begin{equation}
    \hat{G} = \omega_{\rm{RF}} \hat{P}_g\hat{F}_z \hat{P}_g + \omega \hat{P}_e =
    \matrix{\omega_{\rm{RF}}&0&0&0\\
    0&0&0&0\\
    0&0&-\omega_{\rm{RF}}&0\\
    0&0&0&\omega}.
\end{equation}
This leads to the transformation:
\begin{equation}
\begin{split}
    \hat{U}_{\rm{grwa}} & = \hat{P}_g e^{-i \omega_{\rm{RF}} \hat{F}_z t} + \hat{P}_{e}e^{-i \omega t}\\ & =
    \matrix{e^{-i\omega_{\rm{RF}}t}&0&0&0\\
    0&1&0&0\\
    0&0&e^{i\omega_{\rm{RF}}t}&0\\
    0&0&0&e^{-i\omega t}}.
\end{split}
\end{equation}
Applying the above transformation and neglecting rapidly
oscillating terms at frequencies $2\omega_{\rm{RF}}$ and
$2\omega$, the explicit time dependence is eliminated from the
Hamiltonian. This results in effective energy level shifts
described by the diagonal matrix: $\text{diag}\left([
-\omega_{\rm{RF}},0,\omega_{\rm{RF}},-\omega ]\right)$.

\subsection{Alternative forms of the master equation}
\label{app:A4}

It is important to note that different conventions exist for
formulating the quantum master equation---also known as the
quantum Liouville equation, also known as the Lindblad or
Gorini-Kossakowski-Lindblad-Sudarshan (GKLS) equation.  In the
context of this paper, two commonly used forms are particularly
relevant: one prevalent in AMO physics, and another widely adopted
in quantum information and quantum optics.

The most commonly used form of the master equation in quantum
optics is given in Eq.~\eqref{eq:03_master_equation}, where
relaxation processes are described by a set of independent jump
operators $\hat{L}_{\mu}$. This formulation is particularly
convenient, as it allows for a comprehensive characterization of
the system's dissipative dynamics and facilitates further
simplifications using the effective operator or superoperator
formalisms.

In contrast, an alternative yet equivalent form of the master
equation is commonly employed in atomic physics:
\begin{equation}
    \dv{\rho}{t} = \mathcal{L}(\rho) = -i \comm{\hat{H}}{\rho} - \dfrac{1}{2}\acomm{\hat{\Gamma}}{\rho} + \hat{\Lambda}(\rho),
    \label{ME2}
\end{equation}
where $\hat{\Gamma}$ and $\hat{\Lambda}(\rho)$ denote the
relaxation and repopulation operators, respectively
\cite{AuzinshBook, Liang2023}. These operators are usually
introduced phenomenologically, as they capture not only the
internal dynamics of the atomic system but also external
effects---such as atoms leaving the interaction region or being
replenished by new atoms entering the system.

Although the two forms differ in appearance, they are
mathematically equivalent under suitable identifications. Their
structural similarities are readily apparent, enabling a direct
correspondence between the jump-operator and
relaxation-repopulation representations:
\begin{eqnarray}
    \hat{\Gamma} &=& \sum_{\mu} \hat{L}_{\mu}^{\dagger}\hat{L}_{\mu}.\\
    \hat{\Lambda}(\rho) &=& \sum_{\mu} \hat{L}_{\mu} \rho
    \hat{L}_{\mu}^{\dagger}.
\end{eqnarray}
It is worth noting that although the transformation from jump
operators to relaxation and repopulation terms is relatively
straightforward, the inverse transformation is generally ambiguous
and not uniquely defined.

\section{Superoperator formalism}
\label{app:B_superoperators}

As discussed in Sec.~\ref{sec:introduction}, calculating LEPs usually requires expressing the Liouvillian operator in a
superoperator basis. For easier comparison, it is also beneficial
to represent the NHH in the same basis. Since the choice of basis
is not unique, there are multiple possible implementations. Two
common approaches are vectorization into the Fock-Liouville space
using right- and left-hand-side operators, and expansion in the
generalized Gell-Mann basis~\cite{Schlienz1995}.

\subsection{Liouvillians in the generalized Gell-Mann basis}
\label{app:B1}

Here, we recall how the Liouvillian of a $d$-level system can be
represented in the basis formed by the generalized Gell-Mann
matrices, which are also commonly referred to as generalized Pauli
matrices, particularly in the context of quantum information.

To begin, it is important to note that these matrices, together
with the identity matrix---denoted as the set $\lbrace
\hat{\sigma}_i: i<d^2 \rbrace$, where $\hat{\sigma}_{d^2} \propto
\id$---form a basis for real-valued expansions of Hermitian
matrices~\cite{Schlienz1995}. Generally, such a basis can be
constructed following the method outlined in
Ref.~\cite{Schlienz1995}, which reduces to the standard Pauli
matrices for $d=2$ and the standard Gell-Mann matrices for $d=3$.
This property makes the basis particularly well suited for
standard Hermitian Hamiltonians.

For NHHs, however, an additional complication arises: Representing
an arbitrary non-Hermitian $d \times d$ matrix requires extending
the basis. This issue can be addressed by allowing the expansion
coefficients to be complex.

Another crucial property of the generalized Gell-Mann matrices is that they satisfy the standard orthogonality relation, $\Tr{\hat{\sigma}_i \hat{\sigma}_j} = 2 \delta_{ij}.$ This property enables the straightforward vectorization of density matrices as follows:
\begin{equation}
    \hat{\rho} = \sum_{i=1}^{d^2} \left[ \dfrac{1}{2}
    \Tr{\rho \hat{\sigma}_i} \right] \hat{\sigma}_{i} = \sum_{i=1}^{d^2} \rho_i \hat{\sigma}_{i} = \sket{\rho},
    \label{eq:03_rho_GM}
\end{equation}
where $\sket{\rho}$ represents the vectorized density matrix with
elements $\sket{\rho}_i=\Tr{\rho \hat{\sigma}_i}/2$. The next step
is to express the Liouvillian as a linear operator acting on this
vectorized form:
\begin{equation}
    \dhat{ \mathcal{L}}_{ij} = \frac{1}{2}
    \Tr{\mathcal{L}(\hat{\sigma}_j) \hat{\sigma}_i }.
\end{equation}
To achieve this, Eq.~\eqref{eq:03_master_equation} can be
decomposed into three parts: the Hamiltonian component $\dhat{H}$,
the non-Hermitian but coherent part $\dhat{\Gamma}$, and the quantum
jump term $\dhat{\Lambda}$. This separation naturally leads to the
introduction of the NHH superoperator defined as
$\dhat{H}_{\rm{NH}} = \dhat{H} + i \dhat{\Gamma}$:
\begin{equation}
\begin{split}
    \mathcal{L}(\rho) \rightarrow \dhat{\mathcal{L}} \sket{\rho} &= \left( -i \dhat{H} + \dhat{\Gamma} + \dhat{\Lambda} \right) \sket{\rho} \\ &= \left( -i \dhat{H}_{\rm{NH}} + \dhat{\Lambda} \right) \sket{\rho}.
    \label{eq:03_Liouvillian_sup}
\end{split}
\end{equation}
The first term expands as:
\begin{equation}
\begin{split}
    \comm{\hat{H}}{\rho} &= \sum_{i=1}^{d^2} \comm{\hat{H}}{\hat{\sigma}_i}\rho_i = \sum_{i,j=1}^{d^2}  \dfrac{1}{2}
    \Tr{\comm{\hat{H}}{\hat{\sigma}_{i}} \hat{\sigma}_j}
    \rho_i \hat{\sigma}_j\\
    &= \sum_{i,j=1}^{d^2}\hat{\sigma}_{j}\dhat{H}_{ji} \sket{\rho}_i = \dhat{H}\sket{\rho},
\end{split}
\end{equation}
where $\dhat{H}$ is a matrix with elements given by
\begin{equation}
    \dhat{H}_{ij} = \dfrac{1}{2} \Tr{\comm{\hat{H}}{\hat{\sigma}_j}\hat{\sigma}_i}.
    \label{eq:03_sup_H}
\end{equation}
Analogously, the second term in Eq.~\eqref{eq:03_master_equation}
expands as:
\begin{equation}
\begin{split}
    & - \dfrac{1}{2}
    \sum_{\mu}\acomm{\hat{L}_{\mu}^{\dagger}\hat{L}_{\mu}}{\rho}
    \\
   &\quad= - \dfrac{1}{2} \sum_{i,j=1}^{d^2} \left[ \dfrac{1}{2}\sum_{\mu}
   \Tr{ \acomm{\hat{L}_{\mu}^{\dagger}\hat{L}_{\mu}}{\hat{\sigma}_i} \hat{\sigma}_j}
   \right]\rho_{i}\hat{\sigma}_j\\
   &\quad=\sum_{i,j=1}^{d^2} \hat{\sigma}_{j} \dhat{\Gamma}_{ji} \sket{\rho}_{i} = \dhat{\Gamma} \sket{\rho},
\end{split}
\end{equation}
where $\dhat{\Gamma}$ is given by
\begin{equation}
    \dhat{\Gamma}_{ij} = - \dfrac{1}{4} \sum_{\mu}
\Tr{\acomm{\hat{L}_{\mu}^{\dagger}\hat{L}_{\mu}}{\hat{\sigma}_{j}}
\hat{\sigma}_{i} }.
    \label{eq_Gamma_general}
\end{equation}
The last term, describing quantum jumps, reads
\begin{equation}
\begin{split}
    \sum_{\mu} \hat{L}_{\mu}^{\dagger} \rho \hat{L}_{\mu} &= \sum_{i,j=1}^{d^2} \dfrac{1}{2} \sum_{\mu}
\Tr{ \hat{L}_{\mu}^{\dagger}\hat{\sigma}_{i}\hat{L}_{\mu} \hat{\sigma}_{j} } \rho_{i} \hat{\sigma}_{j}\\
    &= \sum_{i,j=1}^{d^2} \hat{\sigma}_{j} \dhat{\Lambda}_{ji} \sket{\rho}_{i} = \dhat{\Lambda} \sket{\rho},
\end{split}
\end{equation}
where $\dhat{\Lambda}$ is given by
\begin{equation}
    \dhat{\Lambda}_{ij} = \dfrac{1}{2} \sum_{\mu}
    \Tr{ \hat{L}_{\mu}^{\dagger}\hat{\sigma}_{j}\hat{L}_{\mu} \hat{\sigma}_{i} }.
    \label{eq_Lambda_general}
\end{equation}

\subsection{Liouvillians in the Liouville-Fock basis}
\label{app:B2}

Expanding a Liouvillian of $d$-level system in the generalized
Gell-Mann basis is a natural choice, especially from the
perspective of experimentally implemented process tomography.
However, this approach can initially seem counterintuitive when
the system of interest is not simply a single-spin system---for
example, a two-level system where the excited state is explicitly
retained rather than eliminated. Despite this, such impressions
are generally misleading, as the generalized Gell-Mann basis can
represent any finite-dimensional system.

Alternatively, this opens the door to a different approach based
on expansion in the Liouville-Fock basis. This method inherently
vectorizes the density matrix~\cite{Abo2024}, transforming it as
follows:
\begin{equation}
    \rho = \sum_{i,j} \rho_{ij} \dyad{i}{j} \quad \longrightarrow \quad \sket{\rho} = \sum_{i,j} \rho_{ij} \ket{i} \otimes \ket{j^*}.
    \label{eq:03_rho_LF}
\end{equation}
In this formalism, the density matrix is transformed into a vector
by sequentially stacking its elements column by column.

To complete the framework, it is necessary to define the
right-hand-side (RHS) and left-hand-side (LHS) acting
superoperators, denoted as $\dhat{R}[\hat{\Gamma}]$ and
$\dhat{L}[\hat{\Gamma}]$, respectively. These superoperators can be
expressed as:
\begin{equation}
\begin{split}
    \dhat{L} \left[ \hat{\Gamma} \right] \sket{\rho}  &=
    \left( \hat{\Gamma} \otimes \id \right) \sket{\rho} = \sum_{i,j} \rho_{ij} \left( \hat{\Gamma} \ket{i} \right) \otimes \ket{j^*}\\
    &\quad \longrightarrow \quad \sum_{i,j} \rho_{ij} \hat{\Gamma} \dyad{i}{j} = \hat{\Gamma}\rho,\\
    \dhat{R} \left[ \hat{\Gamma} \right] \sket{\rho} &=
    \left( \id \otimes \hat{\Gamma}^T \right) \sket{\rho} \quad \longrightarrow \quad \rho \hat{\Gamma}.
\end{split}
\end{equation}
By adopting this convention, the Liouvillian
\eqref{eq:03_master_equation} in the master equation can be
expressed as
\begin{equation}
\begin{split}
    \dhat{\mathcal{L}} &= -i \left( \hat{H} \otimes \id - \id \otimes \hat{H}^{T} \right)\\ &\quad + \sum_{\mu} \left[ \hat{L}_{\mu} \otimes \hat{L}_{\mu}^* - \frac{1}{2} \left( \hat{L}_{\mu}^{\dagger} \hat{L}_{\mu} \otimes \id + \id \otimes \hat{L}_{\mu}^{T} \hat{L}_{\mu}^{*} \right)
    \right].
\end{split}
\end{equation}

\subsection{Properties of the Hamiltonian superoperator}

Following the definition of the superoperator $\dhat{H}_{ij}$ in
Eq.~\eqref{eq:03_sup_H}, we establish two key properties: it is
represented by an antisymmetric and purely imaginary matrix.

First, we demonstrate the antisymmetry property:
\begin{equation}
\begin{split}
    \dhat{H}_{ij} &= \dfrac{1}{2}
    \Tr{\comm{\hat{H}}{\hat{\sigma}_j}\hat{\sigma}_i}\\
    &= \dfrac{1}{2} \left\lbrace\Tr{\hat{H}\hat{\sigma}_j \hat{\sigma}_i}
    - \Tr{\hat{\sigma}_j \hat{H} \hat{\sigma}_i}\right\rbrace\\
    &=\dfrac{1}{2} \left\lbrace\Tr{\hat{H}\hat{\sigma}_j \hat{\sigma}_i}
    - \Tr{\hat{H} \hat{\sigma}_i \hat{\sigma}_j}\right\rbrace = -\dhat{H}_{ji}.
\end{split}
\end{equation}
This shows that $\dhat{H}$ is antisymmetric. Importantly, this
result relies only on the cyclic property of the trace and thus
holds regardless of the chosen basis.

Next, we demonstrate that the elements $\dhat{H}_{ij}$ are purely
imaginary by computing the complex conjugate of its matrix
elements:
\begin{equation}
\begin{split}
    \left( \dhat{H}_{ij} \right)^* &= \dfrac{1}{2}
    \Tr{\comm{\hat{H}}{\hat{\sigma}_j}\hat{\sigma}_i}^* \\
    &= \dfrac{1}{2}
\Tr{ \left\lbrace\comm{\hat{H}}{\hat{\sigma}_j}\hat{\sigma}_i\right\rbrace ^{\dagger}}\\
    &=\dfrac{1}{2}
\Tr{\hat{\sigma}_i^{\dagger}\comm{\hat{H}}{\hat{\sigma}_j}^{\dagger}}.
\end{split}
\end{equation}
Since both $\hat{\sigma}_i$ and $\hat{H}$ are Hermitian, we can
use the property:
\begin{equation}
\begin{split}
    \comm{\hat{H}}{\hat{\sigma}_i}^{\dagger} &= \left( \hat{H}\hat{\sigma}_i - \hat{\sigma}_i\hat{H} \right)^{\dagger}\\
    &= \hat{\sigma}_i^{\dagger}\hat{H}^{\dagger} - \hat{H}^{\dagger}\hat{\sigma}_i^{\dagger} =  \comm{\hat{\sigma}_i}{\hat{H}}.
\end{split}
\end{equation}
Substituting this into our previous expression and applying the
cyclic property of the trace, we obtain:
\begin{equation}
\begin{split}
     \dhat{H}_{ij} ^* &= \dfrac{1}{2}
\Tr{\hat{\sigma}_i^{\dagger}\comm{\hat{H}}{\hat{\sigma}_j}^{\dagger}}
=\dfrac{1}{2} \Tr{\hat{\sigma}_i\comm{\hat{\sigma}_j}{\hat{H}}}\\
     &=\Tr{\comm{\hat{\sigma}_j}{\hat{H}} \hat{\sigma}_i}
     = - \Tr{\comm{\hat{H}}{\hat{\sigma}_j} \hat{\sigma}_i}\\
     &=-\dhat{H}_{ij}.
\end{split}
\end{equation}
This confirms that $\dhat{H}_{ij}$ is purely imaginary.

\subsection{Properties of the non-Hermitian coherent part in
Liouvillians}

Analogously, we now establish the key properties of the
superoperator $\dhat{\Gamma}_{ij}$ defined in
Eq.~\eqref{eq_Gamma_general}. In particular, we show that it is
symmetric and real-valued.

The symmetry property follows by an argument similar to the
antisymmetry of $\dhat{H}$:
\begin{equation}
\begin{split}
    \dhat{\Gamma}_{ij} &= -\dfrac{1}{4}
\Tr{\acomm{\hat{\Gamma}}{\hat{\sigma}_j}\hat{\sigma}_i}\\ &=
-\dfrac{1}{4} \left[ \Tr{\hat{\Gamma}\hat{\sigma}_j
\hat{\sigma}_i}  +
\Tr{\hat{\sigma}_j \hat{\Gamma} \hat{\sigma}_i}\right]\\
    &=-\dfrac{1}{4} \left[\Tr{\hat{\Gamma}\hat{\sigma}_j \hat{\sigma}_i}  +
    \Tr{\hat{\Gamma} \hat{\sigma}_i \hat{\sigma}_j}\right] = \dhat{\Gamma}_{ji}.
\end{split}
\end{equation}
This confirms that $\dhat{\Gamma}$ is a symmetric matrix.

To demonstrate that $\dhat{\Gamma}$ is real-valued, we first note
that although the jump operators $\hat{L}_{\mu}$ may be
non-Hermitian, the operator $\hat{\Gamma} =
\hat{L}_{\mu}^{\dagger}\hat{L}_{\mu}$ is always Hermitian. This
observation allows us to write:
\begin{equation}
    \acomm{\hat{\Gamma}}{\hat{\sigma}_{i}}^{\dagger} = \acomm{\hat{\Gamma}^{\dagger}}{\hat{\sigma}_i^{\dagger}} = \acomm{\hat{\Gamma}}{\hat{\sigma}_i}.
\end{equation}
Using similar reasoning as in the previous case, we compute the
complex conjugate:
\begin{equation}
\begin{split}
      \dhat{\Gamma}_{ij}^* &= -\dfrac{1}{4}
      \Tr{\acomm{\hat{\Gamma}}{\hat{\sigma}_j}\hat{\sigma}_i}^*\\ &= -\dfrac{1}{4}
      \Tr{\hat{\sigma}_i^{\dagger}\acomm{\hat{\Gamma}}{\hat{\sigma}_j}^{\dagger}} \\
     &=-\dfrac{1}{4} \Tr{\hat{\sigma}_i\acomm{\hat{\Gamma}}{\hat{\sigma}_j}}\\ &=
     \Tr{\acomm{\hat{\Gamma}}{\hat{\sigma}_j} \hat{\sigma}_i} =\dhat{\Gamma}_{ij}.
\end{split}
\end{equation}
This confirms that $\dhat{\Gamma}$ is a real-valued matrix.

\subsection{Correspondence between operator and superoperator
spectra} \label{app:D_correspondance}

It is straightforward to observe that a superoperator, represented
by a $d^2 \times d^2$ matrix, has $d^2$ eigenvalues, whereas the
underlying Hamiltonian operator from which it is derived possesses
only $d$ eigenvalues. This discrepancy implies that the
superoperator spectrum must exhibit a degree of redundancy or
degeneracy, as it cannot encode more information than the original
operator's spectrum. Consequently, one expects a direct and
structured relationship between the spectra of the operator and
its associated superoperator.

To investigate this relationship, we first observe that for any
linear operator, the right and left eigenvalue problems yield
identical eigenvalues, even though the corresponding eigenvectors
are generally distinct. This equivalence arises because both
formulations are governed by the same characteristic equation:
\begin{eqnarray}
    \hat{H}_{\rm{NH}}\ket{u} = E \ket{u},
    &\quad& \bra{v}\hat{H}_{\rm{NH}} = E \bra{v}, \nonumber\\
    \left( \hat{H}_{\rm{NH}} - \id E \right) \ket{u} = 0,
    &\quad& \bra{v} \left( \hat{H}_{\rm{NH}} - \id E \right) =
    0.
\end{eqnarray}
Thus the characteristic equation in both cases takes the form:
\begin{equation}
    \det(\hat{H}_{\rm{NH}} - \id E) = 0.
\end{equation}
To analyze the eigenvalue problem of the corresponding
superoperator $\dhat{H}_{\rm{NH}}$, we proceed as follows:
\begin{equation}
    \dhat{H}_{\rm{NH}} \sket{\lambda} = \lambda \sket{\lambda} \quad \Rightarrow \quad \hat{H}_{\rm{NH}} \hat{X} - \hat{X} \hat{H}_{\rm{NH}}^{\dagger} = \lambda \hat{X},
\end{equation}
where the superoperator is expressed in operator form, as given in
Eq.~\eqref{eq:03_HNH_commutator}, and $X$ denotes the operator
whose vectorized form is $\sket{\lambda}$ [see
e.g.,~\eqref{eq:03_rho_GM} or \eqref{eq:03_rho_LF}].

Since $\hat{X}$ must be preserved under the action of
$\hat{H}_{\rm{NH}}$ from both the left and the right, it is
natural to construct $\hat{X}$ from the right and left
eigenvectors of $\hat{H}_{\rm{NH}}$, such that $\hat{X}_{ij} =
\dyad{u_i}{v_j}$. Substituting this form into the eigenvalue
equation yields:
\begin{equation}
\begin{split}
    \dhat{H}_{\rm{NH}} \sket{\lambda_{ij}} &= \hat{H}_{\rm{NH}} \hat{X}_{ij} - \hat{X}_{ij}  \hat{H}_{\rm{NH}}^{\dagger}\\
&= \hat{H}_{\rm{NH}} \dyad{u_i}{v_j} -
\dyad{u_i}{v_j}\hat{H}_{\rm{NH}}^{\dagger} \\&= \left( E_i - E_j^*
\right) \dyad{u_i}{v_j}.
\end{split}
\label{eq:correspondance}
\end{equation}
This derivation demonstrates that the spectrum of the
non-Hermitian superoperator $\dhat{H}_{\rm{NH}}$ is fully
determined by the eigenvalues of the underlying operator
$\hat{H}_{\rm{NH}}$. However, the superoperator spectrum is not
merely a subset of the operator spectrum; instead, it exhibits a
richer structure formed from all possible pairwise differences
between the operator's eigenvalues and their complex conjugates.

The structure of Eq.~\ref{eq:correspondance} reveals that spectral
degeneracies in the NHH operator, i.e., when $E_i = E_j$, directly
translate into degeneracies in the corresponding superoperator
spectrum. Specifically, the eigenvalues $ \lambda_{ii} =
\lambda_{ij} = \lambda_{ji} = \lambda_{jj} = 2
\operatorname{Im}(E_i) $ become degenerate.

This observation can be generalized: an $n$-fold degeneracy in the
spectrum of an NHH operator induces at least an $n^2$-fold
degeneracy in the spectrum of the corresponding superoperator. It
is important to emphasize that this represents only a lower bound
on the degeneracy order in the superoperator spectrum---additional
degeneracies may emerge depending on the specific structure of the
operator and the relationships between its left and right
eigenvectors.

\section{Additional isotropic relaxation in the superoperator description}
\label{app:C_isotropic}

Following the approach outlined in
Sec.~\ref{sec:no_RF_case_operator}, we incorporate isotropic
relaxation within the superoperator framework. To enable a direct
comparison with the results obtained for LEPs, we introduce the
following modified form of the hybrid Liouvillian:
\begin{equation}
    \dhat{\mathcal{L}}_{g}(q) = \dhat{\mathcal{L}}(q) +\dhat{\Gamma}_{\rm{g}} + q \dhat{\Lambda}_{\rm{g}},
\end{equation}
where $\dhat{\mathcal{L}}(q)$ is the hybrid Liouvillian defined in
Eq.~\eqref{eq:hybrid_Liouvillian_matrix}. The operators
$\dhat{\Gamma}_{g}$ and $\dhat{\Lambda}_{g}$ represent,
respectively, the hyperfine relaxation and repopulation (quantum
jump) contributions in the superoperator basis. As a result, the
matrix representation of the modified hybrid Liouvillian becomes
\begin{equation}
    \dhat{\mathcal{L}}_{g}(q) = \dhat{\mathcal{L}}(q) - \gamma \left(
\begin{array}{ccccccccc}
 1 & 0 & 0 & 0 & 0 & 0 & 0 & 0 & 0 \\
 0 & 1 & 0 & 0 & 0 & 0 & 0 & 0 & 0 \\
 0 & 0 & 1 & 0 & 0 & 0 & 0 & 0 & 0 \\
 0 & 0 & 0 & 1 & 0 & 0 & 0 & 0 & 0 \\
 0 & 0 & 0 & 0 & 1 & 0 & 0 & 0 & 0 \\
 0 & 0 & 0 & 0 & 0 & 1 & 0 & 0 & 0 \\
 0 & 0 & 0 & 0 & 0 & 0 & 1 & 0 & 0 \\
 0 & 0 & 0 & 0 & 0 & 0 & 0 & 1 & 0 \\
 0 & 0 & 0 & 0 & 0 & 0 & 0 & 0 & 1-q \\
\end{array}
\right).
\end{equation}
In general, this modification substantially affects the
superoperator spectrum. However, in the limiting cases of $q=0$
(corresponding to the NHH limit) and $q=1$ (representing the pure
Liouvillian case), the resulting spectral changes are notably
simple and analytically tractable.

For the pure NHH limit ($q=0$), the original spectrum [see
Eq.~\eqref{eq:NHH_spectrum}] is modified to:
\begin{eqnarray}
&&\{-\gamma,-2 \Omega-\gamma
,-\alpha_1^{*}-\gamma,-\alpha_1^{*}-\gamma,-\alpha_{1}-\gamma,\nonumber \\
&&-\alpha_{1}-\gamma, -2\alpha_{1}-\gamma,
-2\alpha_{1}^{*}-\gamma,-2\Omega-\gamma \},
\end{eqnarray}
which corresponds to a uniform shift of the real part of the
spectrum by $-\gamma$. This aligns with the nature of isotropic
relaxation, which imposes uniform damping on all eigenmodes.

Similarly, in the $q=1$ case (the fully trace-preserving
Liouvillian regime), all eigenvalues are uniformly shifted by
$-\gamma$, with the exception of the stationary state at $\lambda
= 0$. This behavior highlights the isotropic nature of the
relaxation process while ensuring the trace preservation of the
density matrix.

\section*{Acknowledgements}

We gratefully acknowledge insightful discussions with Szymon
Pustelny, Yujie Sun, and Arash Dezhang Fard. This work was
supported by the Polish National Science Centre (NCN) under the
Maestro Grant No. DEC-2019/34/A/ST2/00081.


\end{document}